\documentclass[12pt,preprint]{aastex631}

\usepackage{mathtools}

\submitjournal{ApJ}

\shorttitle{High $\Sigma_1$ Blue Spirals}
\shortauthors{Hao et al.}

\begin{document}

\title{Are High $\Sigma_1$ Massive Blue Spiral Galaxies Rejuvenated Systems?}

\correspondingauthor{Cai-Na Hao}
\email{hcn@bao.ac.cn}

\author[0000-0002-0901-9328]{Cai-Na Hao}
\affiliation{Tianjin Astrophysics Center, Tianjin Normal University, Tianjin 300387, China}

\author{Xiaoyang Xia}
\affiliation{Tianjin Astrophysics Center, Tianjin Normal University, Tianjin 300387, China}

\author[0000-0002-8614-6275]{Yong Shi}
\affiliation{School of Astronomy and Space Science, Nanjing University, Nanjing 210093, China}
\affiliation{Key Laboratory of Modern Astronomy and Astrophysics (Nanjing University), Ministry of Education, Nanjing 210093, China}

\author[0000-0002-3275-9914]{Rui Guo}
\affiliation{Tianjin Astrophysics Center, Tianjin Normal University, Tianjin 300387, China}

\author[0000-0003-3226-031X]{Yanmei Chen}
\affiliation{School of Astronomy and Space Science, Nanjing University, Nanjing 210093, China}
\affiliation{Key Laboratory of Modern Astronomy and Astrophysics (Nanjing University), Ministry of Education, Nanjing 210093, China}

\author[0000-0002-9767-9237]{Shuai Feng}
\affiliation{College of Physics, Hebei Normal University, 20 South Erhuan Road, Shijiazhuang 050024, China}
\affiliation{Hebei Key Laboratory of Photophysics Research and Application, Shijiazhuang 050024, China}

\author[0000-0002-1971-5458]{Junqiang Ge}
\affiliation{National Astronomical Observatories, Chinese Academy of Sciences, Beijing 100101, China}

\author[0000-0002-3890-3729]{Qiusheng Gu}
\affiliation{School of Astronomy and Space Science, Nanjing University, Nanjing 210093, China}
\affiliation{Key Laboratory of Modern Astronomy and Astrophysics (Nanjing University), Ministry of Education, Nanjing 210093, China}

\begin{abstract} 
	
Quiescent galaxies generally possess denser cores than star-forming galaxies
	with similar mass. As a measurement of the core density, the central
	stellar mass surface density within a radius of 1\,kpc ($\Sigma_1$) was
	thus suggested to be closely related to galaxy quenching. Massive
	star-forming galaxies with high $\Sigma_1$ do not fit into this
	picture. To understand the origin of such galaxies, we compare the
	spatially-resolved stellar population and star formation properties of
	massive ($ > 10^{10.5}{\rm M}_{\odot}$) blue spiral galaxies with high
	and low $\Sigma_1$, divided by $\Sigma_1 = 10^{9.4} M_\odot \, {\rm
	kpc}^{-2}$, based on the final release of MaNGA IFU data. We find that
	both high $\Sigma_1$ and low $\Sigma_1$ blue spirals show large
	diversities in stellar population and star formation properties. Despite
	the diversities, high $\Sigma_1$ blue spirals are
	statistically different from the low $\Sigma_1$ ones. Specifically, the
	radial profiles of the luminosity-weighted age and Mgb/${\rm \langle Fe
	\rangle}$ show that high $\Sigma_1$ blue spirals consist of a larger
	fraction of galaxies with younger and less $\alpha$-element enhanced
	centers than their low $\Sigma_1$ counterparts, $\sim 55\%$ versus
	$\sim 30\%$. The galaxies with younger centers mostly have higher
	central specific star formation rates, which still follow the
	spaxel-based star formation main sequence relation though. Examinations
	of the H$\alpha$ velocity field and the optical structures suggest that
	galactic bars or galaxy interactions should be responsible for the
	rejuvenation of these galaxies. The remaining $\sim 45\% $ of high
	$\Sigma_1$ blue spirals are consistent with the inside-out
	growth scenario.
  
\end{abstract}




\section{INTRODUCTION} \label{sec:intro} 

How a galaxy quenches is a key question to understand in the context of galaxy
evolution. Both observational and theoretical studies suggest that the buildup
of a central dense core is a requisite for galaxy quenching
\citep[e.g.,][]{Cheung2012, Fang2013, Zolotov2015, Barro2017, Dekel2019}.
Nowadays, a widely used probe for a central dense core is the stellar mass
surface density within a radius of 1\,kpc, denoted as $\Sigma_1$.  The
connection between the presence of a dense core and galaxy quenching is
manifested by the tight correlation between $\Sigma_1$ and the total stellar
mass for quiescent galaxies \citep{Cheung2012, Fang2013}. A similar
correlation also exists for star-forming galaxies, but it shows a steeper
slope, a smaller normalization and a much larger scatter. Such correlations
have existed since $z\sim3$, and hence it was suggested that star-forming
galaxies evolve along the relations \citep{Barro2017}. As soon as they
experienced a phase of significant core growth, known as compaction process,
which led to a rapid increase in $\Sigma_1$ and depletion of cold gas, they
would migrate up to the relation of quiescent population and start quenching
\citep{Zolotov2015, Barro2017, Dekel2019}. Therefore, galaxies with high $\Sigma_1$
are expected to be quenched systems. 

Interestingly, it was noted that some massive ($ > 10^{10.5}{\rm M}_{\odot}$)
blue galaxies lie on the relation for quiescent galaxies \citep{Fang2013,
Guo2020}, named high $\Sigma_1$ blue spirals hereafter. They have dense cores
but are still forming stars. Although they only account for a small fraction of
blue spirals ($\sim 10\%$) \citep{Guo2020}, they are important for our
understanding of the whole picture of galaxy formation and evolution. 
\citet{Fang2013} suggested that high $\Sigma_1$ blue spirals are candidates for
rejuvenation, and they pointed out that the rejuvenated star formation mainly
carries on in the outer parts.  The more recent study by \citet{Guo2020} found
that high $\Sigma_1$ blue spirals possess massive bulges, large host dark
matter halo mass and high bar/ring or interacting/merger incidence rate,
similar to massive red (i.e., passive) spirals. On the basis of these results,
the authors conjectured that high $\Sigma_1$ blue spirals may be rejuvenated
red spirals.  The rejuvenation scenario is consistent with studies based on
cosmological simulations \citep{Tacchella2016} and observations
\citep{Tacchella2022}, which revealed that star-forming galaxies oscillate
about the star formation main sequence relation due to the gas depletion and
replenishment. A galaxy that had obtained its dense core via compaction process
and quenched would form stars again as long as new gas fed it. The
galaxy with re-ignited star formation would appear as a high $\Sigma_1$ blue
spiral, which is a natural result in such a process.

However, \citet{Woo2019} proposed another possibility. Based on the
survey of the Mapping Nearby Galaxies at the Apache Point Observatory (MaNGA)
\citep{Bundy2015} integral field unit (IFU) spectra, they found that the
centers of galaxies with higher $\Sigma_1$ have younger age, enhanced specific
star formation rate (sSFR) and less metals compared to their outer parts, which
is in well agreement with the compaction-like core-building scenario. The
mechanisms for core-building are less clear though. 

In this work, we aim to deepen our understanding of the possible formation
channels for high $\Sigma_1$ blue spirals by focusing on their
spatially-resolved stellar population and star formation properties derived
from MaNGA data. Low $\Sigma_1$ blue spirals will be used as a comparison sample.
Throughout this paper, a flat $\Lambda$CDM
cosmology with $\Omega_{\rm m}=0.3$, $\Omega_{\rm \Lambda}=0.7$, $H_{\rm
0}=70\,{\rm km \, s^{-1} Mpc^{-1}}$ is adopted. 

\section{SAMPLE AND PARAMETERS} \label{sec:sample}

\begin{figure}
\plotone{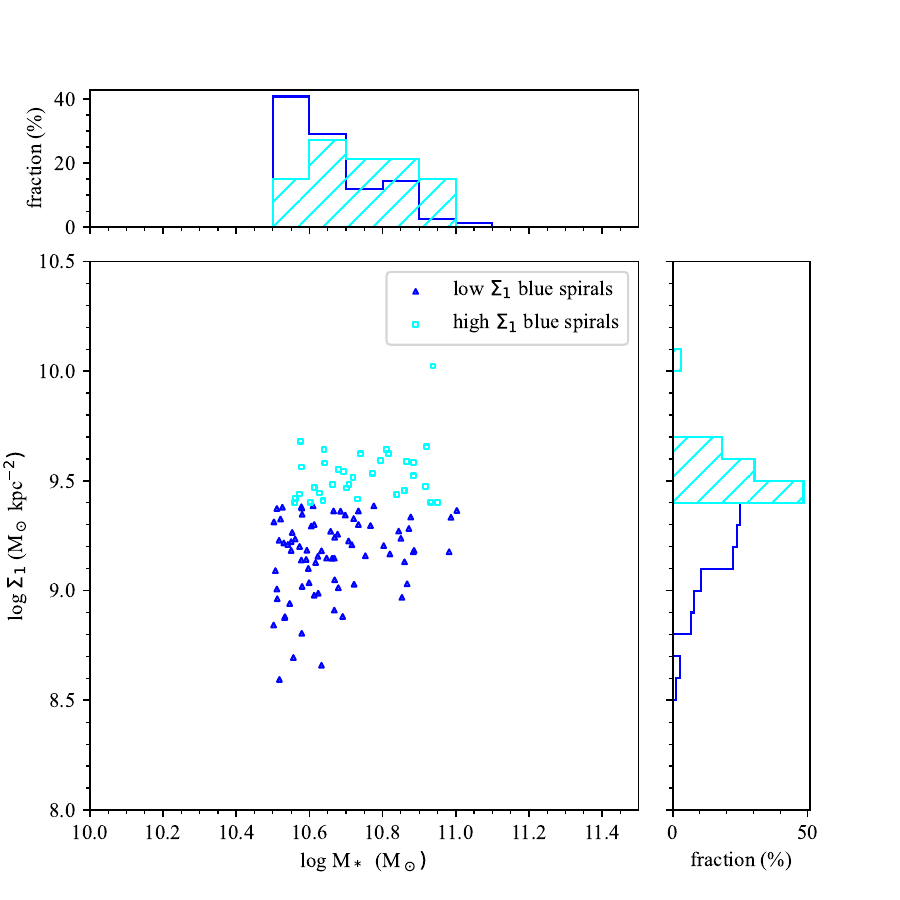}
	\caption{\label{stellarmass_Sigma1.pdf} Stellar mass and $\Sigma_1$ distributions for
	our sample of blue spirals. Low and high $\Sigma_1$ blue spirals are represented by
	blue triangles and cyan squares, respectively.
	}
\end{figure}

The samples used in this work are similar to the sample of blue spirals in
\citet{Hao2019}, but updated with the final data release of the MaNGA survey,
which is part of the 17th data release of the Sloan Digital Sky Survey (SDSS)
\citep{Abdurrouf2022}. They consist of 109 blue spirals within a stellar mass
range of $10^{10.5} - 10^{11}M_\odot$, under the assumption of a Chabrier
initial mass function (IMF) \citep{Chabrier2003}, and a redshift range of $0.02
- 0.05$. These galaxies were drawn from a parent sample in
\citet{Mendel2014} with $0.02 < z < 0.05$, which are included in the Legacy
area of the SDSS DR7. The morphological information was retrieved from the
Galaxy Zoo 1 \citep{Lintott2008,Lintott2011}, and they were separated in color
by their locations in the dust-corrected $u-r$ color vs.  stellar mass diagram.
Specifically, $(u-r)_{\rm corr} < -0.583+0.227\,{\rm log}(M_{\rm
*}/M_\odot)$ was used to single out the blue galaxies. The readers can find
more details on sample selection in \citet{Hao2019} and \citet{Guo2020}. In
this work, to further distinguish blue spirals with dense cores from those with
less dense cores, we use $\Sigma_1 = 10^{9.4} M_\odot \, {\rm kpc}^{-2}$ as a
dividing point\footnote{In \citet{Guo2020}, the best-fit to the red spirals
produced a relation of log $\Sigma_1 = (0.48\pm0.06)$ log $(M_*/M_\odot) +
(4.20\pm0.60)$. For the median of the stellar mass range of $10^{10.5} -
10^{11}M_\odot$, the corresponding $\Sigma_1$ is $10^{9.4} M_\odot \, {\rm
kpc}^{-2}$.}. This yields 33 high $\Sigma_1$ and 76 low $\Sigma_1$ blue
spirals. The stellar mass and $\Sigma_1$ distributions for the two types of
blue spirals are shown in Figure~\ref{stellarmass_Sigma1.pdf}. To
evaluate the reasonability of our adopted threshold for separating high
$\Sigma_1$ from low $\Sigma_1$ galaxies, we examined our results by dividing the blue spirals
evenly into three subsamples, which correspond to a log $\Sigma_1$ range of $<
9.2$, $9.2-9.4$ and $> 9.4$. Interestingly, the statistical properties of
galaxies within the middle bin are also in between those of the first and last
bins but are much more similar to the low $\Sigma_1$ bin in most cases.
Therefore, the results based on the samples selected using our adopted dividing
point should be robust.

To investigate the global star formation properties of our sample
galaxies, we drew the total stellar masses and star formation rates (SFRs) from
\citet{Mendel2014} and \citet{Salim2018}, respectively. Both parameters were
derived using Spectral Energy Distribution (SED) fitting. The stellar masses
are based on the SDSS $u, g, r, i, z$ five bands photometry, while the SFRs are
based on the combined UV to IR photometry from GALEX, SDSS and WISE. Both
stellar masses and SFRs were derived under the assumption of a Chabrier IMF.

For the purpose of investigating possible formation channels of high $\Sigma_1$
blue spirals, we derive azimuthally averaged radial profiles of stellar
population and star formation properties, and the kinematic asymmetry of the
gas content from the two-dimensional maps. To measure the azimuthally averaged
radial profiles for each individual galaxy, the $r-$band elliptical Petrosian
50\% light radius ($R_e$), the axis ratio (b/a) and position angle (PA) used
for elliptical apertures were extracted from the NASA-Sloan Atlas (NSA) catalog
\citep{Blanton2011}. A radial bins size of 0.15\,$R_{\rm e}$ was adopted.

Following \citet{Hao2019}, we derived stellar population properties, measured
by mass-weighted and luminosity-weighted age and metallicity, from the MaNGA
Pipe3D value added catalog (VAC) for DR17 \citep{Sanchez2022}. Compared to DR15
\citep{Sanchez2016, Sanchez2018}, the authors adopted an improved version of
the fitting code \citep{Lacerda2022} and a new SSP spectral library based on
the MaNGA stellar library \citep[MaStar;][]{Yan2019} in DR17. Both versions of
the Pipe3D VACs adopted a Salpeter IMF \citep{Salpeter1955}. For common objects
in the 15th and 17th data release, we carefully compared the results from these
two versions and found that not only the absolute values but also the radial
profiles of the stellar population properties of blue spirals changed from DR15
to DR17. Only the luminosity-weighted age is almost immune to the changes in
different versions of Pipe3D VAC. Therefore, we only use the
luminosity-weighted age and the model-independent parameter Mgb/${\rm \langle
Fe \rangle}$ (=Mgb/(0.5*Fe5270+0.5*Fe5335)) to probe the stellar populations in
this paper.  Mgb/${\rm \langle Fe \rangle}$ is an $\alpha$-element enhancement
indicator, which is widely used to measure the star formation timescale. The
measurements of the metal absorption lines were taken from the MaNGA data
analysis pipeline (DAP) \citep{Westfall2019}. By examination, we found that
there are 11 high $\Sigma_1$ and 24 low $\Sigma_1$ blue spirals without
reliable Fe measurements because of the contamination of strong night sky
emission lines. Hence, they will not be included in the analysis of Mgb/${\rm
\langle Fe \rangle}$ profiles.

To study the status of the interstellar medium (ISM) of our sample
galaxies, we distinguished star-forming regions from composites and active
galactic nuclei (AGNs) in the [OIII]/H$\beta$ versus [NII]/H$\alpha$ BPT
diagram \citep{Baldwin1981} developed by \citet{Kauffmann2003} and
\citet{Kewley2001} on a spaxel basis. For the measurement of the SFR, we used
the [OIII]/H$\beta$ versus [SII]/H$\alpha$ diagram calibrated by
\citet{Kewley2006} to single out star-forming regions. SFRs based on the
[SII]/H$\alpha$-identified star-forming spaxels are similar to those based on a
combination of star-forming regions and composites selected by the
[OIII]/H$\beta$ versus [NII]/H$\alpha$ diagram. The Milky Way dust
extinction-corrected emission line fluxes were drawn from the MaNGA DAP
\citep{Belfiore2019, Westfall2019}. Only the emission lines with
signal-to-noise ratio (S/N) greater than 3 were used. We corrected the
H$\alpha$ emission for internal dust attenuation based on the Balmer decrement
H$\alpha$/H$\beta$ and the O’Donnell \citep{Donnell1994} reddening law. An
intrinsic value of 2.86 was adopted for H$\alpha$/H$\beta$.  For spaxels with
H$\alpha$/H$\beta$ below the intrinsic value, no dust attenuation correction
was applied. The SFRs were calculated from the dust-corrected H$\alpha$
luminosities, using a conversion factor of $8.77\times{10^{-42}}$ calibrated
using STARBURST99 \citep{Leitherer1999, Vazquez2005, Leitherer2010,
Leitherer2014} and under the assumption of a Salpeter IMF, following
\citet{Kennicutt2009}. To obtain the radial profile of sSFR ($=SFR/M_\ast$), we
drew the stellar mass surface density from the Pipe3D VAC, and used the ratio
of the total SFR to the total stellar mass contained within each elliptical
annulus as a measure of the sSFR at the corresponding radius. For the
spaxel-based deprojected stellar mass and SFR surface density, we followed
\citet{Barrera-Ballesteros2016} to correct for the inclination effect ($\Sigma
= \Sigma_{\rm obs} \times (b/a)$). We note that we adopted a Chabrier IMF for
the sample selection and the study of the global properties inheriting from our
previous work, and assumed a Salpeter IMF for the analyses of the
spatially-resolved stellar population and star formation properties for the
sake of using the Pipe3D VAC. But this will not affect our results and
conclusions, considering that only a constant conversion factor needs to be
applied to convert the stellar mass and SFR from one IMF to the other.

The kinematic asymmetry was obtained by fitting the velocity field of the
ionized gas using the KINEMETRY package \citep{Krajnovic2006}, following
\citet{Feng2020}. Specifically, the average kinematic asymmetry within
1\,$R_{\rm e}$ ($\bar{v}_{\rm asym}$) was used to characterize the entire
galaxy. Because of the low S/N of their H$\alpha$ emission lines, we cannot
obtain reliable estimates of $\bar{v}_{\rm asym}$ for 10 high $\Sigma_1$ and 17
low $\Sigma_1$ blue spirals, as judged by their larger relative uncertainties,
$\delta\bar{v}_{\rm asym} > 0.01$ \citep{Feng2022}. Therefore, they were
removed from the sample in the statistics of $\bar{v}_{\rm asym}$.

\section{RESULTS} \label{sec:results}

\begin{figure}
\plottwo{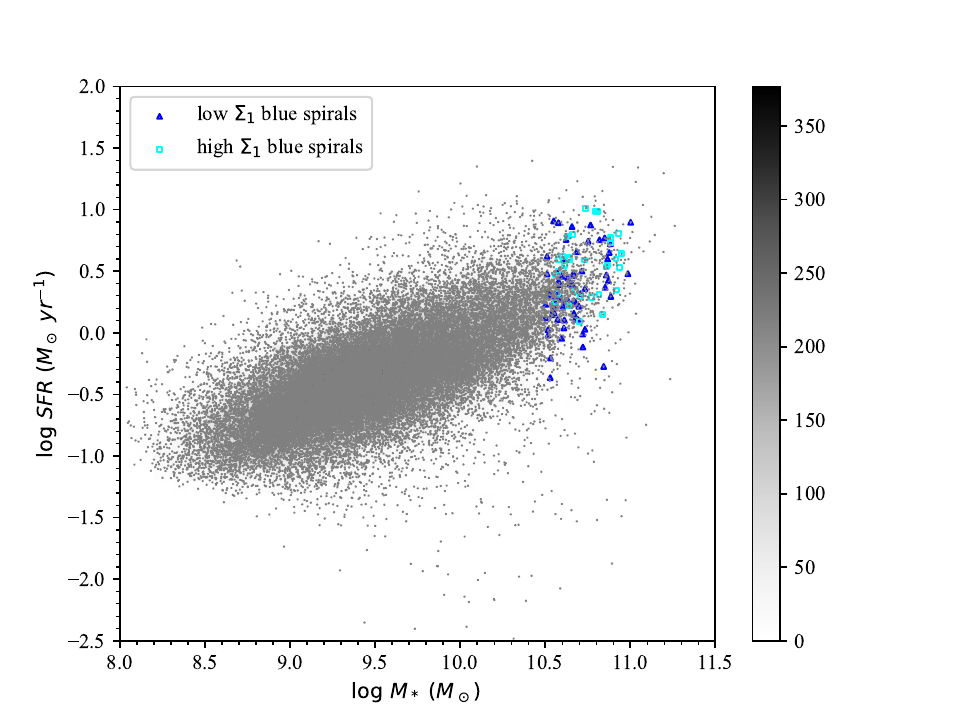}{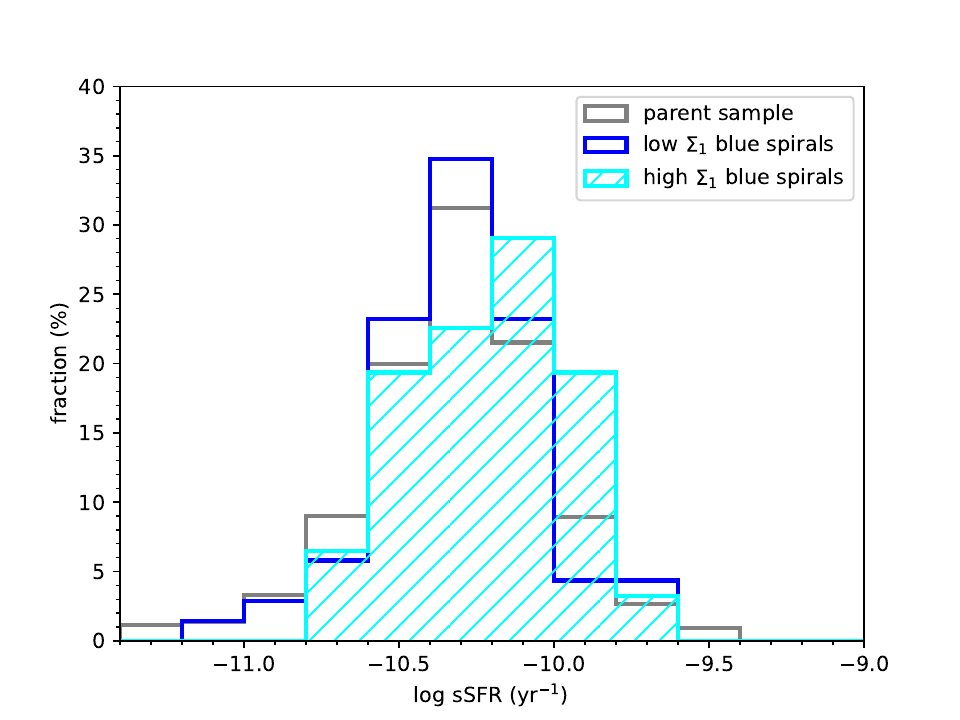}
	\caption{\label{SF_totalquantity.pdf} The global star formation
	properties of the two samples of blue spirals in terms of the star
	formation main sequence relation ({\it left}) and the distribution of
	the specific star formation rate ({\it right}). In the left panel, the
	two-dimensional histogram and the grey data points represent the parent
	sample with star-forming centers, while the blue triangles and cyan squares represent the low and
	high $\Sigma_1$ blue spirals, respectively. In the right panel, the
	colors of the histograms have the same meaning as those in the left
	panel, but the grey histogram only includes galaxies with $10^{10.5} <
	M_{\rm *} < 10^{11}M_\odot$ in the parent sample for a fair
	comparison.}

\end{figure}

Before diving into the spatially-resolved properties, we first examine the
global star formation properties of high and low $\Sigma_1$ blue spirals. The
left panel of Figure~\ref{SF_totalquantity.pdf} shows the locations of the blue
spirals and their parent sample in the SFR versus $M_{\rm *}$ diagram. To
better define the star formation main sequence, we only include galaxies
with star-forming centers in the parent sample in
Figure~\ref{SF_totalquantity.pdf}, as classified using their central 3\arcsec\, diameter SDSS fiber spectra.  We do not fit a relation to the data
points, since the best-fit depends both on the sample selection and on the
fitting method. Instead, we look at the distributions directly. It is obvious
that both low and high $\Sigma_1$ blue spirals occupy the massive end of the
distribution of their parent sample, and they do not deviate from the sequence.
This is further confirmed by the right panel of
Figure~\ref{SF_totalquantity.pdf}, which shows the histograms of sSFR for the
two types of blue spirals and the parent sample with star-forming centers and $10^{10.5}
< M_{\rm *} < 10^{11}M_\odot$. Although the high $\Sigma_1$ blue spirals
consist of a larger fraction of galaxies with higher sSFR than the low
$\Sigma_1$ blue spirals and the parent sample, they are not too different from
them, as indicated by the Kolmogorov-Smirnov (KS) test. The probability that
high and low $\Sigma_1$ blue spirals are drawn from the same distribution is
3.5\%, while the KS test yields a probability of 7.9\% for high $\Sigma_1$ blue
spirals and the parent sample. The standard deviations of sSFR for high
$\Sigma_1$, low $\Sigma_1$ blue spirals and the parent sample are 0.25, 0.27
and 0.27, respectively.

To shed light on the possible origins of high $\Sigma_1$ blue spirals, we need to look into the
spatially-resolved properties.

\subsection{Stellar Populations} \label{subsec:stepop}

\begin{figure}
\plotone{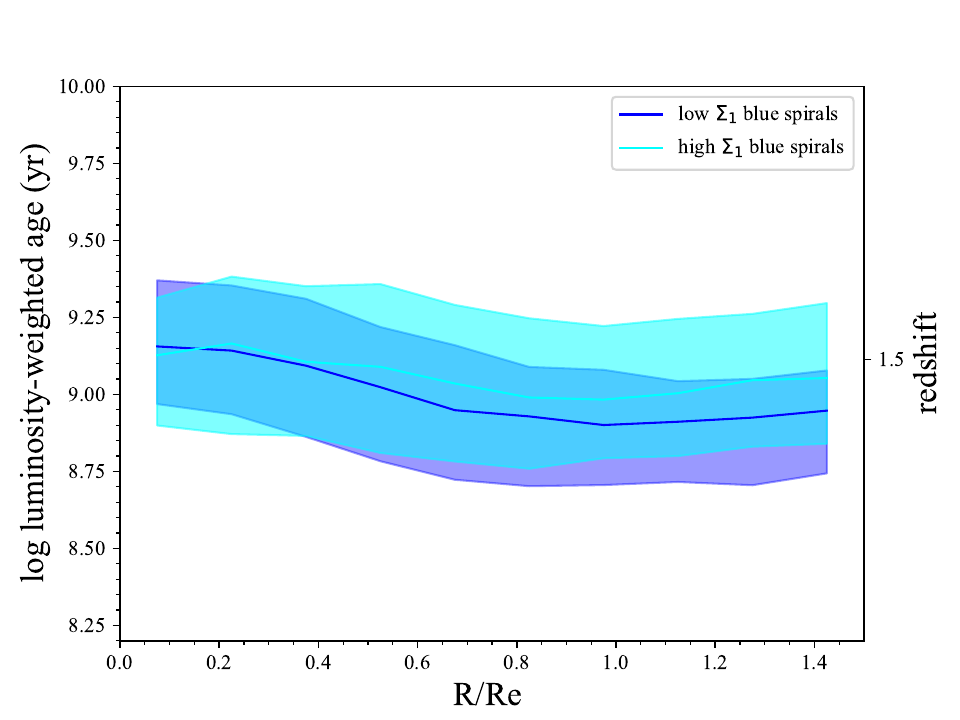}
	\caption{\label{lwage_prof.pdf} Radial profiles of
	luminosity-weighted age for high $\Sigma_1$ (cyan) and low
	$\Sigma_1$ (blue) blue spirals, respectively. The solid
	lines show the median of the respective samples, and the shaded regions
	represent the 16\% and 84\% of the distributions.}
\end{figure}

\begin{figure}
\plotone{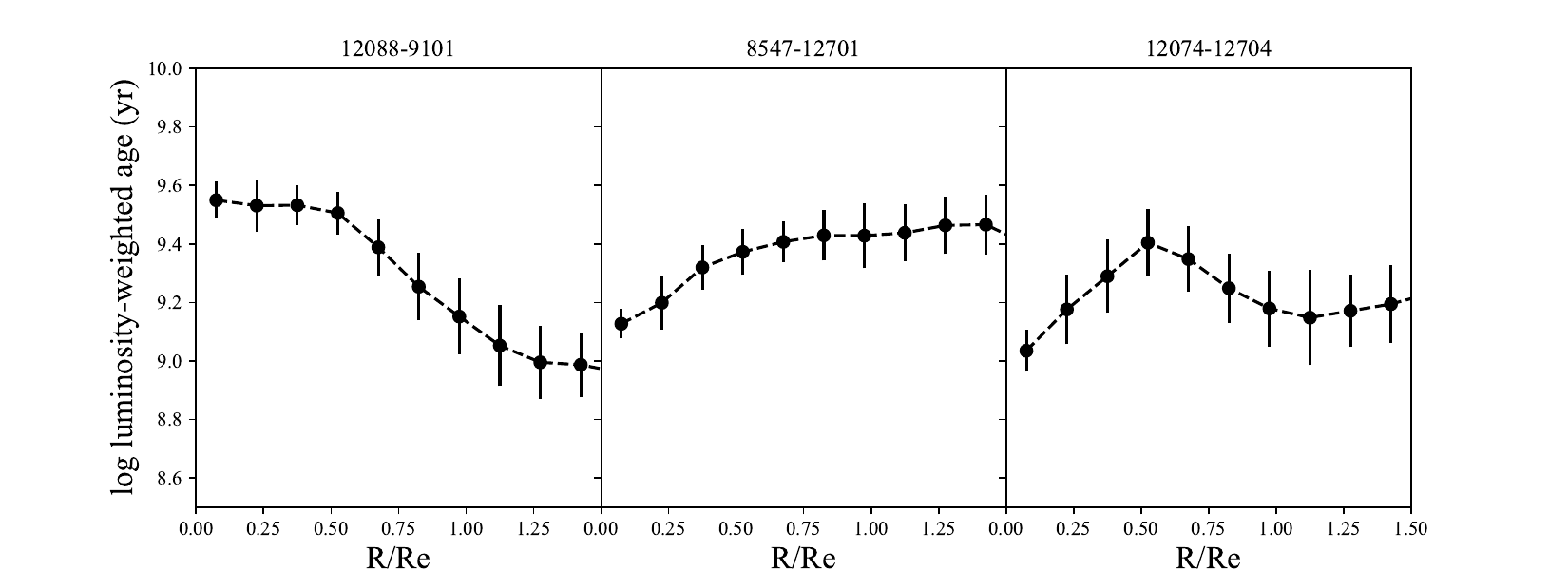}
        \caption{\label{lwage_radprof_highsigma1BS_example.eps}
        Luminosity-weighted age profiles for three examples of high $\Sigma_1$
        blue spirals. The solid circles and the associated error bars are the
        median and standard deviation of the data within each elliptical annulus. From left
        to right, the three panels show three types of age profiles that are
        representative of blue spirals.}
\end{figure}

We first investigate the luminosity-weighted age, which is sensitive to
newly formed stars. Adopting the widely used method in studies using IFU data,
we plot the median radial profiles and their 16\% and 84\% percentiles of the
luminosity-weighted age for the two types of galaxies in
Figure~\ref{lwage_prof.pdf}. Both the median and the 16\% and 84\% percentiles
indicate that the two types of blue spirals are similar in the inner age
profiles, but high $\Sigma_1$ blue spirals show older stellar populations than
low $\Sigma_1$ blue spirals towards the outer part of galaxies. More
strikingly, both high and low $\Sigma_1$ blue spirals show broad age
distribution across the probed radius, as indicated by the spread of the
shadows in Figure~\ref{lwage_prof.pdf}. This suggests a large diversity in the
age profile within each population. Therefore, we examine the age profiles one
by one to both understand the whole population and reveal the true difference
between high and low $\Sigma_1$ blue spirals. It turns out that the age
profiles can be roughly divided into three categories for both types of blue
spirals, as illustrated by the three examples in
Figure~\ref{lwage_radprof_highsigma1BS_example.eps}. From left to right, the
panels show a declining profile, an increasing profile, and a profile with an
age peak at some intermediate radius, respectively. The galaxies with the third
type of age profiles are actually the ``turnover'' galaxies, as identified and
named by \citet{LinLin2017}. They pointed out that the turnover feature is
closely linked with the bar structure \citep{LinLin2017,LinLin2020}. We also
find that the majority of our ``turnover'' galaxies are barred. Evidently,
galaxies with the latter two kinds of profiles, i.e., increasing and
``turnover'' radial profiles, have relatively younger centers and older outer
parts. For high $\Sigma_1$ blue spirals, the three types of profiles account
for $45.5^{+8.7}_{-8.2}\%$ ($15^{+3}_{-3}$/33), $30.3^{+9.0}_{-6.7}$\%
($10^{+3}_{-2}$/33) and $24.2^{+8.8}_{-5.9}$\% ($8^{+3}_{-2}$/33) of the
sample. In comparison, the corresponding percentages are $71.0^{+4.6}_{-5.7}$\%
($54^{+4}_{-4}$/76), $14.5^{+5.0}_{-3.1}$\% ($11^{+4}_{-2}$/76) and
$14.5^{+5.0}_{-3.1}$\% ($11^{+4}_{-2}$/76) for the sample of low $\Sigma_1$
blue spirals. The errors represent the 1$\sigma$ binomial confidence limits,
derived using the method of \citet{Cameron2011}. It is obvious that low
$\Sigma_1$ blue spirals consist of a larger fraction of objects with older
centers than high $\Sigma_1$ galaxies. 

\begin{figure}
\plotone{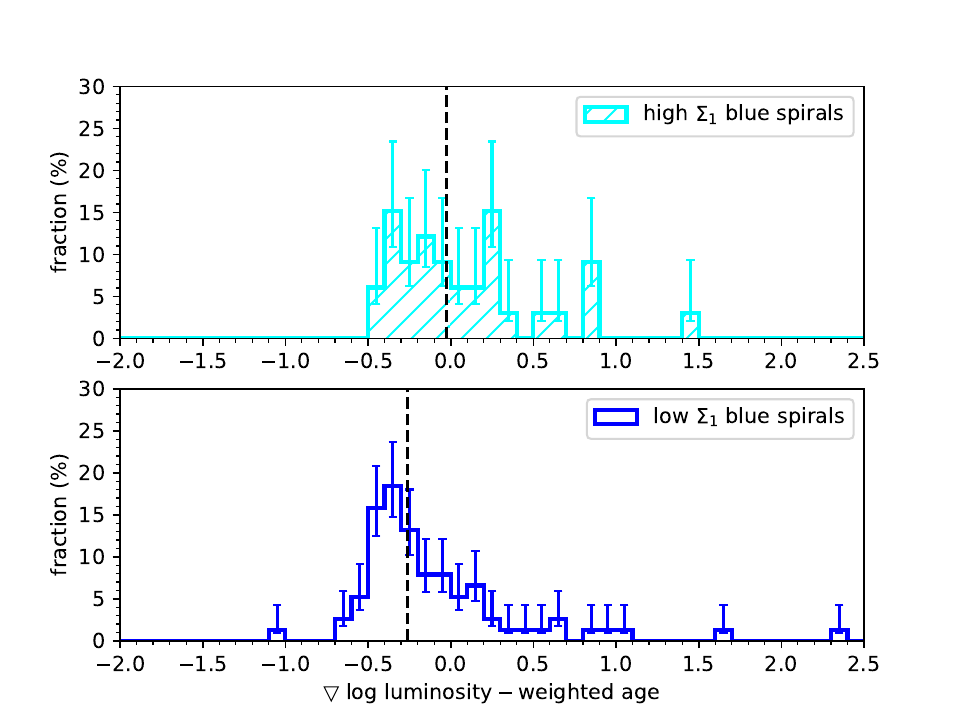}
	\caption{\label{lwagemedianprofile_innergrad.pdf} Distributions of the radial gradient
	of luminosity-weighted age for high $\Sigma_1$ ({\it top}) and low $\Sigma_1$ 
	({\it bottom}) blue spirals, respectively. The error bars represent the 1$\sigma$ binomial confidence limits based on the method of \citet{Cameron2011}.
	}
\end{figure}

Gradients of physical parameters are often used to quantify their radial
profiles. In general, the gradients are derived by fitting a straight line to
the radial profile of the parameter. In the case of luminosity-weighted age, it
would be done by a linear fit to the logarithm age versus radius. However, as
can be seen from Figure~\ref{lwage_radprof_highsigma1BS_example.eps}, a linear
model is not a proper representation of the data here, especially when
considering the whole radius range, which is essential for our purpose of
identifying possible rejuvenated galaxies. By examining the individual age
profiles, we find that a linear fit to the entire radial profile till 1.5\,Re
for the increasing and decreasing profiles and a linear fit to the inner
0.5\,Re profile for the ``turnover'' galaxies can separate the galaxies with
old centers from those with young centers in most cases. Hence, we perform
error-weighted regressions for each individual age profile. During the
fitting, the standard deviations of the luminosity-weighted age within the
elliptical annuli are taken as the errors of the age. A fit to the inner
0.6\,Re profile for the ``turnover'' galaxies does not change the sign of the
gradient. Figure~\ref{lwagemedianprofile_innergrad.pdf} shows the derived age
gradients. It is clear that high $\Sigma_1$ blue spirals tend to
show more positive slopes than the low $\Sigma_1$ blue spirals, which is
resulted from the larger fraction of high $\Sigma_1$ blue spirals with younger
centers compared to low $\Sigma_1$ ones.
According to the age gradients, the number of galaxies with negative (positive)
inner profile is 17 (16) for high $\Sigma_1$ and 55 (21) for low $\Sigma_1$
blue spirals, respectively. They agree with the visual classification based on
the whole radial profiles presented above, confirming the population difference
between high and low $\Sigma_1$ blue spirals.

\begin{figure}
	\plotone{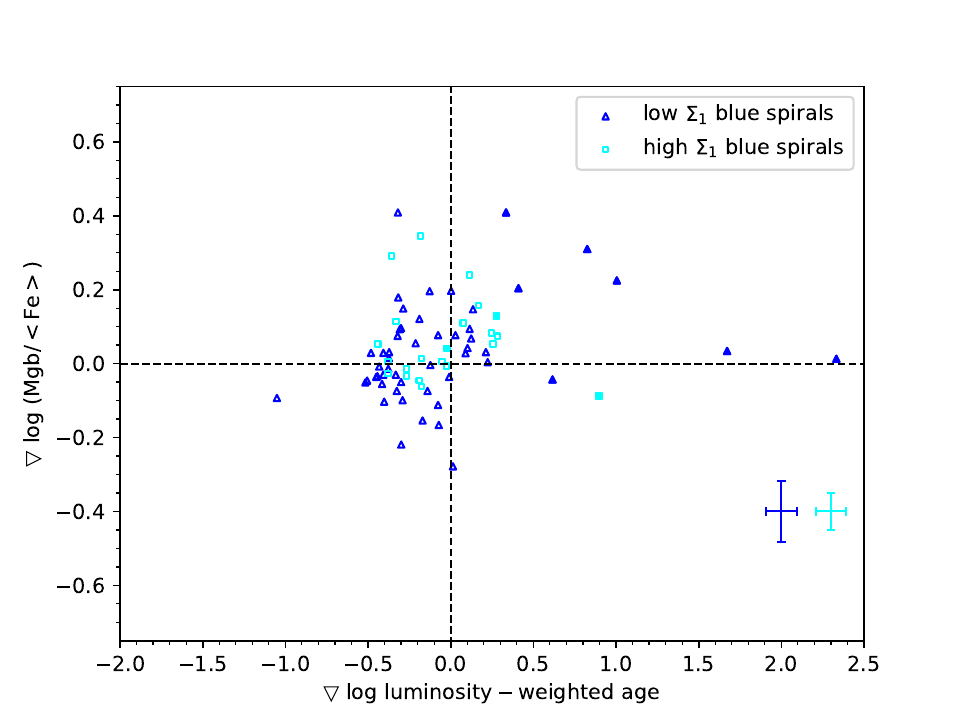}
	\caption{\label{mgbfe_grad_lwage_grad.pdf} Mgb/${\rm \langle Fe
	\rangle}$ gradient as a function of luminosity-weighted age
	gradient. Low and high $\Sigma_1$ blue spirals are represented by blue
	triangles and cyan squares, respectively. The horizontal and vertical
	dashed lines separate galaxies with negative gradients from those with
	positive gradients. Galaxies with ``turnover'' age profiles are
	denoted with filled symbols. The error bars shown in the bottom-right
	corner represent the median measurement errors for low $\Sigma_1$
	(blue) and high $\Sigma_1$ (cyan) blue spirals, respectively.  }
\end{figure}

The model-independent parameter Mgb/${\rm \langle Fe \rangle}$ is a powerful
probe to the star formation timescale \citep{Thomas2005}. It is generally
believed that the earlier a galaxy formed, the shorter the star formation
timescale was. A shorter (longer) star formation timescale leads to a larger
(smaller) Mgb/${\rm \langle Fe \rangle}$. Therefore, a positive correlation
between age and Mgb/${\rm \langle Fe \rangle}$ is expected. Because of the
large scatters, we opt not to show the radial profiles of Mgb/${\rm \langle Fe
\rangle}$. Instead, we use the same method to derive the gradients of the
Mgb/${\rm \langle Fe \rangle}$ profiles as for the age gradients.
Figure~\ref{mgbfe_grad_lwage_grad.pdf} plots the gradient of Mgb/${\rm \langle
Fe \rangle}$ profile as a function of the gradient of luminosity-weighted age.
The horizontal and vertical dashed lines separate galaxies with negative slopes
from those with positive slopes. It is obvious that the one-to-one
correspondence between old (young) age and high (low) Mgb/${\rm \langle Fe
\rangle}$ is not perfect, with the consistency percentages of $\sim ~60\%$
(13/22) and $\sim 70\%$ (36/52) for high and low $\Sigma_1$ blue spirals,
respectively. Interestingly, the remaining $\sim 40\%$ of the high $\Sigma_1$
blue spirals and $\sim 30\%$ of the low $\Sigma_1$ blue spirals without the
age--Mgb/${\rm \langle Fe \rangle}$ correspondence mostly include galaxies that
host old centers with low Mgb/${\rm \langle Fe \rangle}$, i.e., those with
negative age gradient and positive Mgb/${\rm \langle Fe \rangle}$ gradient. In
other words, for almost all galaxies with younger centers, their central
Mgb/${\rm \langle Fe \rangle}$ is low. This may also apply to the galaxies
without reliable Fe measurements. The expected fractions of high and
low $\Sigma_1$ blue spirals with positive Mgb/${\rm \langle Fe \rangle}$
gradient would be about $16_{-3}^{+3}/33$ and $21_{-3}^{+4}/76$, respectively, as
deduced from the age gradients.
The association between positive age gradient and positive Mgb/${\rm \langle Fe
\rangle}$ gradient implies that there are new star formation in the inner
regions of these galaxies. The newly formed stars from the iron-enriched gas by
type Ia supernova could lower the Mgb/${\rm \langle Fe \rangle}$.
Alternatively, if the galaxies have experienced multiple rejuvenations, the
lower Mgb/${\rm \langle Fe \rangle}$ could have been produced by previous
rejuvenations, which took place late enough to allow the type Ia supernova
explosion. The galaxies with old and low Mgb/${\rm \langle Fe \rangle}$ centers
probably had experienced rejuvenations at earlier times.

\subsection{Star Formation Properties}

By selection, both high $\Sigma_1$ and low $\Sigma_1$ blue spirals belong
to the ``blue cloud'' in the dust-corrected $u-r$ versus $M_{\rm *}$ diagram,
and they show similar global sSFR distributions, as shown in
Figure~\ref{SF_totalquantity.pdf}. In this subsection, we study their ISM
status and evaluate the star formation intensity in a spatially-resolved way.

\begin{figure}
	\plottwo{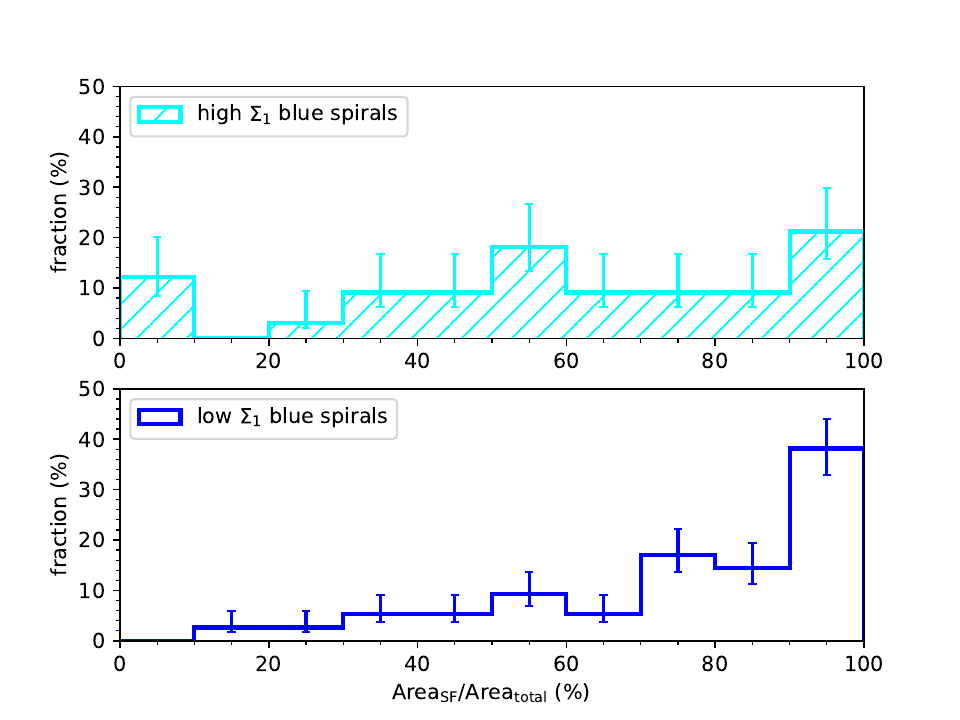}{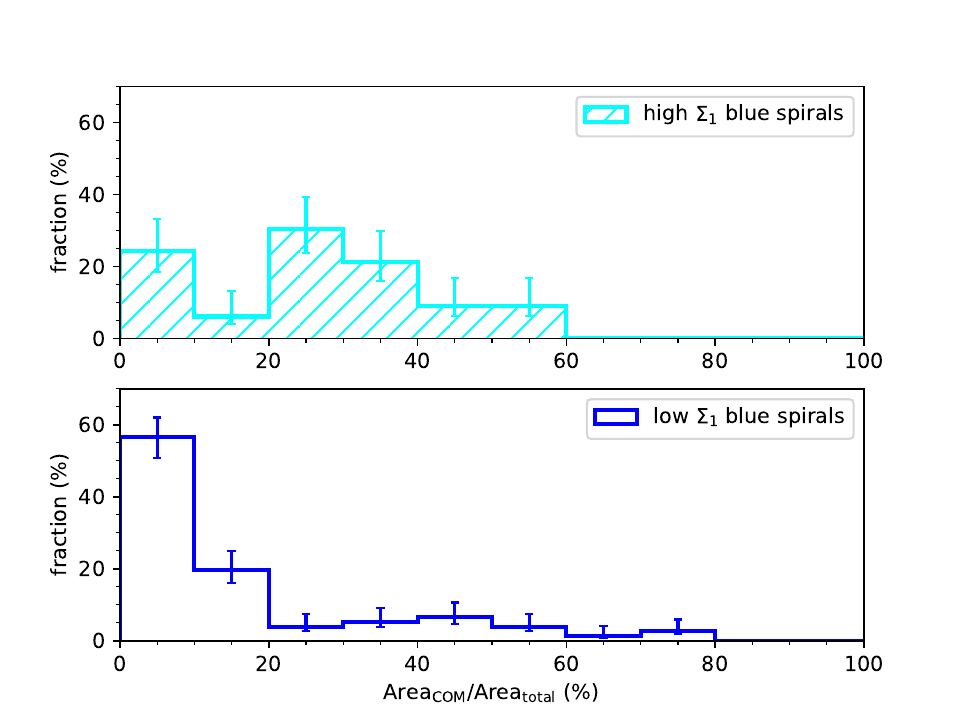}
	\caption{\label{sfareafrac.pdf} Histograms of the total fractional area
	of star formation spaxels ({\it left}) and spaxels with ``composite''
	spectral features ({\it right}) for the high $\Sigma_1$ ({\it top}) and
	low $\Sigma_1$ ({\it bottom}) blue spirals. The error bars represent
	the 1$\sigma$ binomial confidence limits, following the method of
	\citet{Cameron2011}.}
\end{figure}

\begin{figure}
	\plotone{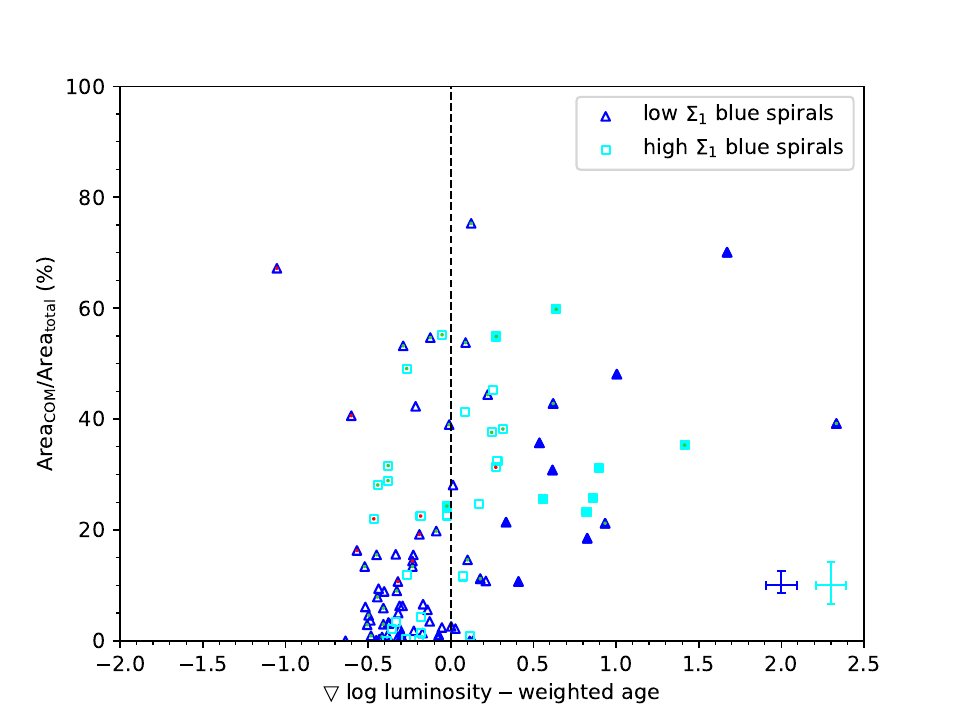}
	\caption{\label{COMareafrac_vs_agegrad.pdf} Total fractional area of
	spaxels with ``composite'' spectral features as a function of
	luminosity-weighted age gradient. The blue triangles and cyan squares
	represent low and high $\Sigma_1$ blue spirals, respectively. The
	vertical dashed line separates galaxies with positive age gradients
	from those with negative age gradients. Galaxies with ``turnover'' age
	profiles are denoted with filled symbols. Small red and green dots in
	the middle of the symbols indicate galaxies hosting central AGN and
	composite spectral features, respectively. The error bars in the
	bottom-right corner are the median measurement errors for low
	$\Sigma_1$ (blue) and high $\Sigma_1$ (cyan) blue spirals.}
\end{figure}

We use the fractional area of star formation spaxels for our samples of
blue spirals to quantify their ISM status. In the left panel of
Figure~\ref{sfareafrac.pdf}, we present the histogram distributions of the
total fractional area of star formation spaxels for the two samples of blue
spirals. It is clear that more than 70\% of low $\Sigma_1$ blue spirals are
forming stars throughout most ($> 70\%$) areas of the galaxies, whereas high
$\Sigma_1$ blue spirals show very extended and relatively flat distribution. We
find that the spaxels that are not identified as star-forming regions are
mostly composite regions except for a small fraction of spaxels with low S/N.
The total areal fraction of composition regions is quantified and plotted in
the right panel of Figure~\ref{sfareafrac.pdf}. It shows clearly that compared
to low $\Sigma_1$ blue spirals, a much larger fraction of high $\Sigma_1$ blue
spirals have more than 20\% and higher fractions of areas dominated by
composite regions. To explore whether there is any association between the age
profile and the ISM status, we plot the age gradients versus the areal fraction
of the composite regions in Figure~\ref{COMareafrac_vs_agegrad.pdf}.
``Turnover'' galaxies were pointed out by solid points. As stated above, most
of them are barred galaxies.  Generally, composite regions are either in galaxy
centers or at the interface between central AGNs and outer star-forming
regions. To estimate the influence of such cases, we label galaxies with
composite or AGN centers using little green and red dots, respectively. It is
interesting to see that most galaxies with positive age gradients, i.e., young
centers, have larger fractions of composite regions regardless of the central
compactness of blue spirals, and more than half of them do not have a composite
or AGN center.  By contrast, for the galaxies with old centers, i.e., negative
age gradients, only a small fraction of low $\Sigma_1$ blue spirals and about
half of the high $\Sigma_1$ blue spirals have larger fractions of composite
regions, and they mostly occur in galaxies with composite or AGN centers.
Therefore, apart from the general cases, composite regions are mainly
associated with barred galaxies and galaxies with positive age gradients. The
two-dimensional BPT diagrams tell that composite regions mostly appear around
bar regions in barred galaxies or in the outer regions of galaxies with rising
age profiles. This suggests that composite regions are linked with old stellar
populations, and the star formation is more concentrated towards the central
regions. In comparison with low $\Sigma_1$ blue spirals, the higher fraction
of high $\Sigma_1$ blue spirals with larger areal fraction of composite regions
are a result of its greater fraction of members with younger centers.

\begin{figure}
	\plotone{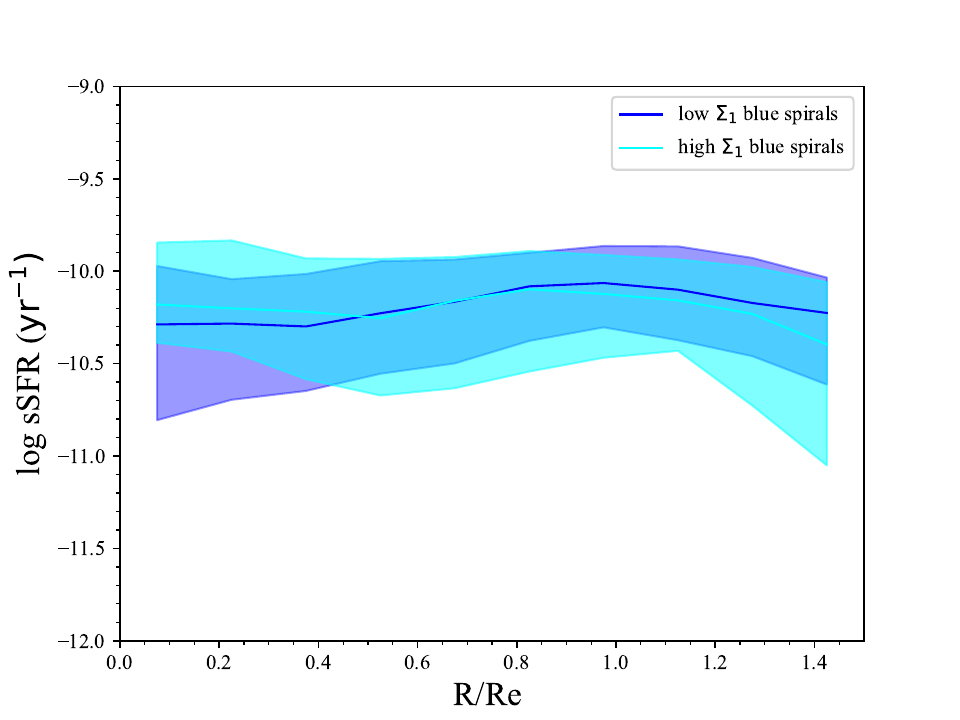}
	\caption{\label{sSFR_prof.pdf} Radial profiles of specific star
	formation rate for high $\Sigma_1$ (cyan) and low $\Sigma_1$ blue
	spirals (blue).  The sSFR within each elliptical annulus was derived
	from the total star formation rate and total stellar mass within the
	respective elliptical annulus. The solid curves and the shaded regions
	indicate the median, and the 16\% and 84\% of the distributions,
	respectively.}
\end{figure}

\begin{figure}
	\plotone{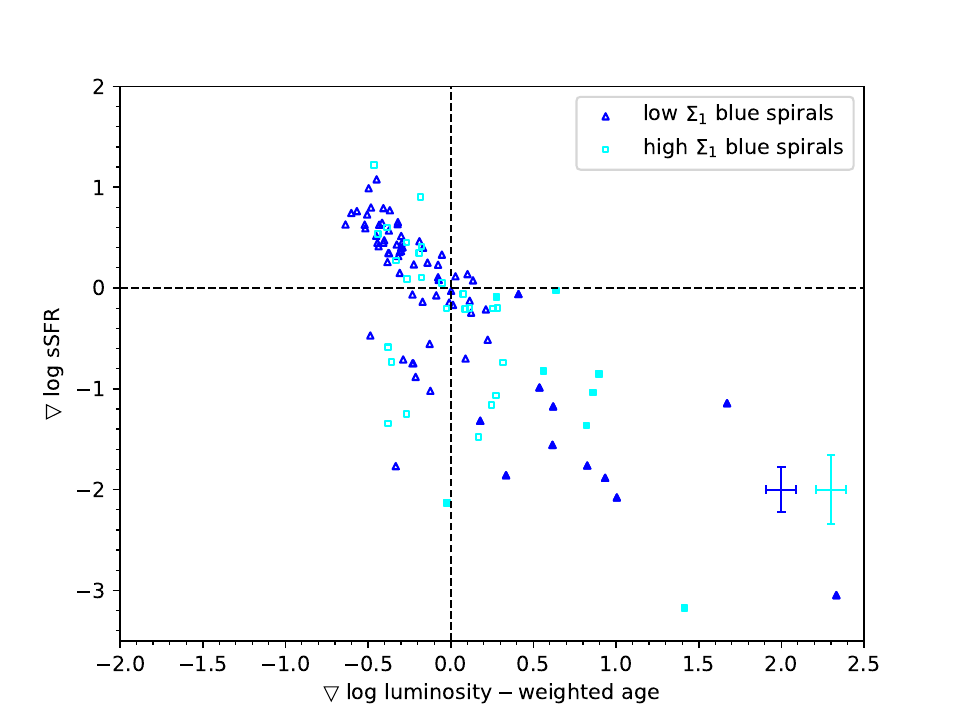}
	\caption{\label{sSFRradprof_vs_lwagemedianprof_grad.pdf} Specific SFR
	gradient as a function of luminosity-weighted age gradient. Low and
	high $\Sigma_1$ blue spirals are represented by blue triangles and cyan
	squares, respectively. The horizontal and vertical dashed lines
	separate galaxies with negative gradients from those with positive
	gradients. Galaxies with “turnover” age profiles are denoted with
	filled symbols. The error bars shown in the bottom-right corner
	represent the median measurement errors for low $\Sigma_1$ (blue) and
	high $\Sigma_1$ (cyan) blue spirals, respectively.  }
\end{figure}

We adopt the most commonly used quantity sSFR to evaluate the star
formation activity, which is also a simple tracer of the star formation
history. From the lessons that we learned from the radial profiles of the
luminosity-weighted age and Mgb/${\rm \langle Fe \rangle}$, we expect a large
diversity in the sSFR profile as well. Nonetheless, to gauge the overall star
formation intensity, we plot the 50\%, 16\% and 84\% of the sSFR profiles for
both types of blue spirals in Figure~\ref{sSFR_prof.pdf}. It can be seen that
for both types of galaxies, the sSFR is less than $\sim 10^{-10} yr^{-1}$,
which indicates that they are not forming stars in a burst mode. A one-by-one
examination for the sSFR and luminosity-weighted age profiles shows that they
have very good correspondence in the sense that a higher sSFR usually
corresponds to a younger age.  Corresponding to the three types of age profiles
shown in Figure~\ref{lwage_radprof_highsigma1BS_example.eps}, there are also
three types of sSFR profiles, i.e., a profile with low sSFR in the center, a
profile with high sSFR in the center, and a profile with a sSFR valley at some
intermediate radius (i.e., ``turnover'' galaxies). Specifically, the
percentages of galaxies with these types of sSFR profiles are
$61.8_{-5.8}^{+5.2}\%$, $7.9_{-2.1}^{+4.3}\%$ and $30.3_{-4.7}^{+5.7}\%$ for
low $\Sigma_1$ blue spirals, and $33.3_{-7.1}^{+9.0}\%$, $30.3_{-6.7}^{+9.0}\%$
and $36.4_{-7.4}^{+8.9}\%$ for high $\Sigma_1$ blue spirals, respectively.
These fractions are not exactly the same as those derived from the age profiles
because the correspondence between age and sSFR profiles is not perfect, 
which is also demonstrated by the anti-correlation between age gradients and sSFR
gradients presented in Figure~\ref{sSFRradprof_vs_lwagemedianprof_grad.pdf}.
For high $\Sigma_1$ blue spirals, $\sim 88\%$ (29/33) of the sample galaxies
show consistent age and sSFR profiles, and for low $\Sigma_1$ blue spirals, the
fraction is $\sim 79\%$ (60/76).

We note that in the measurement of the sSFR profile, the sSFR at each radius is
represented by the ratio of the total SFR to the total stellar mass within the
corresponding elliptical annulus. In essence, it is an average over the
elliptical annulus. Therefore, the sSFR profiles shown in
Figure~\ref{sSFR_prof.pdf} represent an overall star formation property across
the galaxies, which probably conceal some regions with starburst features.
Therefore, we derive the spatially resolved relation between the SFR surface
density and the stellar mass surface density to quantify the star formation
intensity on each spaxel.

\begin{figure}
        \plotone{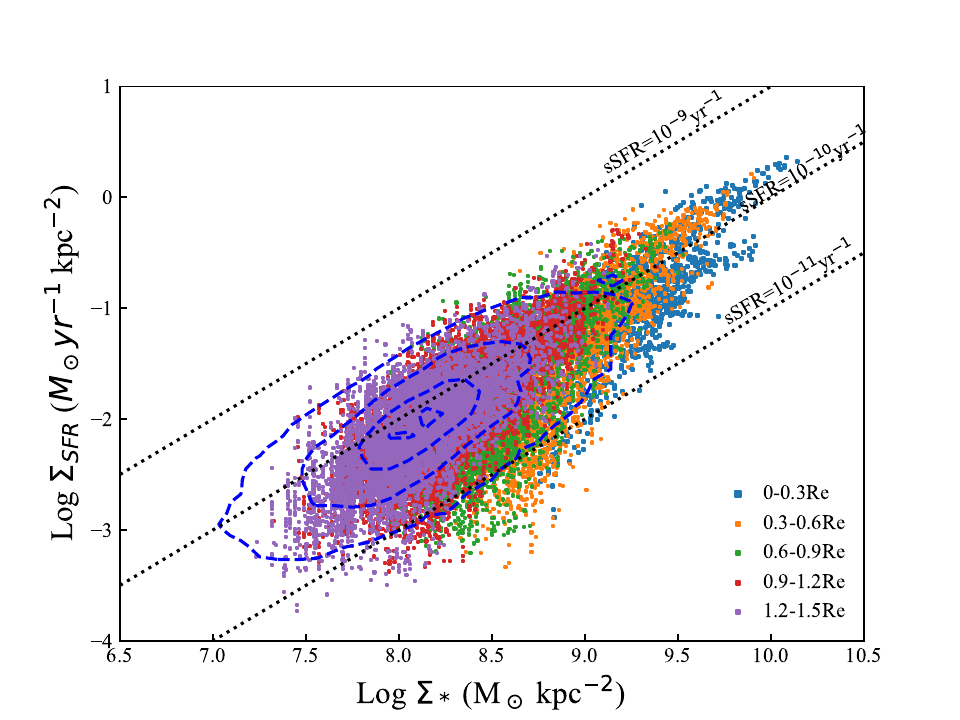}
\caption{\label{resolvedSFMS.pdf} Spatially-resolved star formation main
	sequence relation for [SII]/H$\alpha$-selected star-forming spaxels.
	The blue contours indicate the distribution of low $\Sigma_1$ blue
	spirals. They enclose 5\%, 30\%, 68\% and 95\% of the sample,
	respectively. The high $\Sigma_1$ blue spirals are represented by
	small filled circles color coded by the locations of the spaxels. The
	three dotted lines indicate constant sSFR of $10^{-9}$yr$^{-1}$,
	$10^{-10}$yr$^{-1}$ and $10^{-11}$yr$^{-1}$ respectively, as labeled
	above the corresponding lines.}
\end{figure}

Figure~\ref{resolvedSFMS.pdf} shows the spatially-resolved star formation main
sequence relations, i.e., the relations between the stellar mass surface
density and the SFR surface density. The contours represent the
distribution of the star-forming spaxels in low $\Sigma_1$ blue spirals. The
outermost contour includes 95\% of the spaxels. The star-forming spaxels of
high $\Sigma_1$ blue spirals are represented by the data points color-coded by
their spatial locations in radius. It is obvious that the majority of the
spaxels in high $\Sigma_1$ blue spirals follow a similar distribution to the
low $\Sigma_1$ blue spirals except the high density end. The color-coded data
points show a stratified distribution with the inner spaxels hosting higher
stellar mass surface density and SFR surface density. In spite of the higher
values of SFR surface densities in the inner regions, the data points do not
deviate from the constant sSFR of $10^{-10}$yr$^{-1}$, which roughly serves as
a valid estimate for the low $\Sigma_1$ blue spirals represented by the
contours. Therefore, the spatially-resolved measurements of the star formation
intensity provide a consistent result with the azimuthally averaged measures
shown in Figure~\ref{sSFR_prof.pdf} in the sense that the star formation mode
in high $\Sigma_1$ blue spirals is not significantly different from low
$\Sigma_1$ blue spirals. 

\section{DISCUSSION}\label{sec:discussion}

In this paper, we compare the stellar population and star formation properties
of high $\Sigma_1$ blue spirals with those of low $\Sigma_1$ blue spirals.
The purpose is to understand the origin of the high $\Sigma_1$
blue spirals. As mentioned in Section~\ref{sec:intro}, \citet{Guo2020} found
that high $\Sigma_1$ blue spirals are similar to red spirals in many aspects,
e.g., bulge to total mass ratios, dark matter halo masses and detailed
morphological features. Accordingly, they suggested that high $\Sigma_1$ blue
spirals are rejuvenated systems from quenched red spirals. \citet{Fang2013}
also proposed a rejuvenation origin for high $\Sigma_1$ blue spirals, and they
speculated that the rejuvenated star formation mostly occurs in the outer
regions. On the other hand, \citet{Woo2019} suggested that a compaction-like
process may contribute to the formation of the dense core. 

The results presented in Section~\ref{sec:results} reveal that both low and
high $\Sigma_1$ blue spirals show large diversities in stellar populations and
star formation properties. But they can be broadly divided into two types: one
type includes galaxies with decreasing age profiles, i.e., old centers, and the
other comprises galaxies with younger centers. Compared to low $\Sigma_1$ blue
spirals, high $\Sigma_1$ blue spirals consist of a larger fraction of galaxies
with a younger, less $\alpha$-element enhanced center, and an older, more
$\alpha$-element enhanced disk ($\sim 55\%$ vs. $\sim 29\%$). Furthermore, such
galaxies generally show higher sSFR in the central regions and lower sSFR in
the outer disks. However, both the azimuthally averaged sSFR profiles and the
spatially resolved star formation main sequence relation show that these
central star formation activities are not starbursts. The relatively low sSFRs
($\sim 10^{-10}{\rm yr}^{-1}$) may imply that the current star formation
activities are not responsible for the buildup of the central dense cores,
which formed at an earlier epoch. A rejuvenation scenario is able to explain
the properties of these high $\Sigma_1$ blue spirals with younger centers. In
such a scenario, new gas inflow and star formation triggered by some mechanisms
take place in the quenched galaxies that formed via a fast process at an early
epoch. The gas inflow and star formation have been carrying on across the
galaxies but are more concentrated on the central regions, which lowered the
luminosity-weighted age and Mgb/${\rm \langle Fe \rangle}$ of the stellar
populations.

We next look for possible triggering mechanisms to verify the rejuvenation
scenario. Rejuvenation must be associated with some physical processes that
can cause gas inflow and re-ignition of star formation. It is straightforward
to search for signs for such processes using asymmetry in morphologies and gas
velocity fields.

\begin{figure}
\plotone{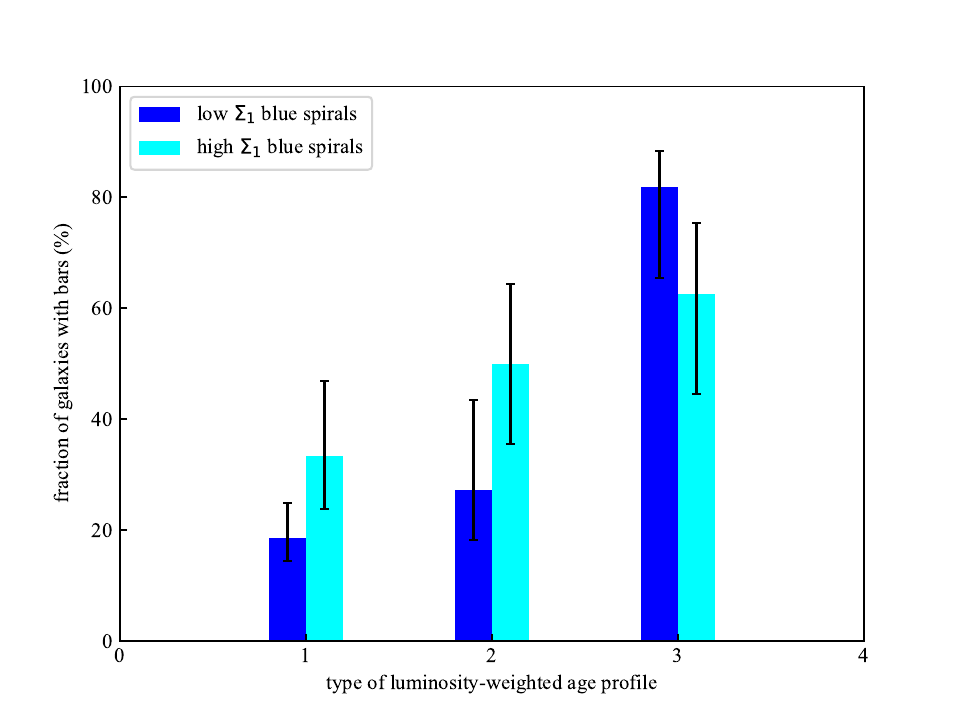}
	\caption{\label{GZ2bar_dis_lwagecode.pdf} Fraction of barred galaxies
	as a function of the type of the luminosity-weighted age profiles for
	low $\Sigma_1$ (blue) and high $\Sigma_1$ (cyan) blue spirals. The
	x-axis represents the type of the luminosity-weighted age profiles,
	number coded by 1, 2 and 3.  One stands for a descending age profile, 2
	stands for an increasing profile, and 3 stands for the age profile with
	a peak (i.e., ``turnover'' feature). The error bars represent the
	1$\sigma$ binomial confidence limits based on the method of
	\citet{Cameron2011}.}
\end{figure}

We retrieve morphological information from the Galaxy Zoo 2 (GZ2) 
\citep{Willett2013,Hart2016}, and adopt the suggested mean vote fraction of 0.5
as the threshold \citep{Hart2016}. We also obtain the information on bar and
tidal features from the MaNGA visual morphology catalog (Vazquez-Mata et al. in
prep).
This catalogue contains a visual morphological classification based on
inspection of image mosaics created using a new digital processing of SDSS and
Dark Energy Legacy Survey (DESI) \citep{Dey2019} images. This new processing
enables the identification of internal structures and low surface brightness
features. After comparing with the GZ2 results, we find that for the
classification of barred galaxies, if we adopt 0.75 as the threshold for
Vazquez-Mata et al., i.e., count galaxies with clear conspicuous bars, the two
catalogs produce consistent results. Furthermore, the identification of tidal
features in Vazquez-Mata et al. is also in good agreement with the GZ2
classification for disturbed morphologies. Since GZ2 includes a much larger
sample than the MaNGA visual morphology catalog, we adopt the GZ2 results for
the following analysis to facilitate future comparison studies. For our samples
of galaxies, the bar fractions for high and low $\Sigma_1$ blue spirals are
$\sim 50\%$ and $\sim 30\%$, respectively. The fractions of galaxies with tidal
features are similar in the two types of galaxies, $\sim 15\%$. These fractions
are more or less consistent with those based on the parent samples, as
presented in \citet{Guo2020}. The association between bar structures and age
profile types is shown in Figure~\ref{GZ2bar_dis_lwagecode.pdf}. It is clearly
seen that galaxies with a ``turnover'' age profile are mostly barred galaxies
for both types of spirals, which was also mentioned in
Section~\ref{subsec:stepop}. Half of the high $\Sigma_1$ blue spirals with an
increasing age profile are unbarred, though. We examined these galaxies and
found that they show either tidal features or rings\footnote{One galaxy is
barred but the mean vote fraction in GZ2 is 0.4, smaller than the threshold of
0.5.}, which are probably the causes for the rejuvenation.

\begin{figure}
\plotone{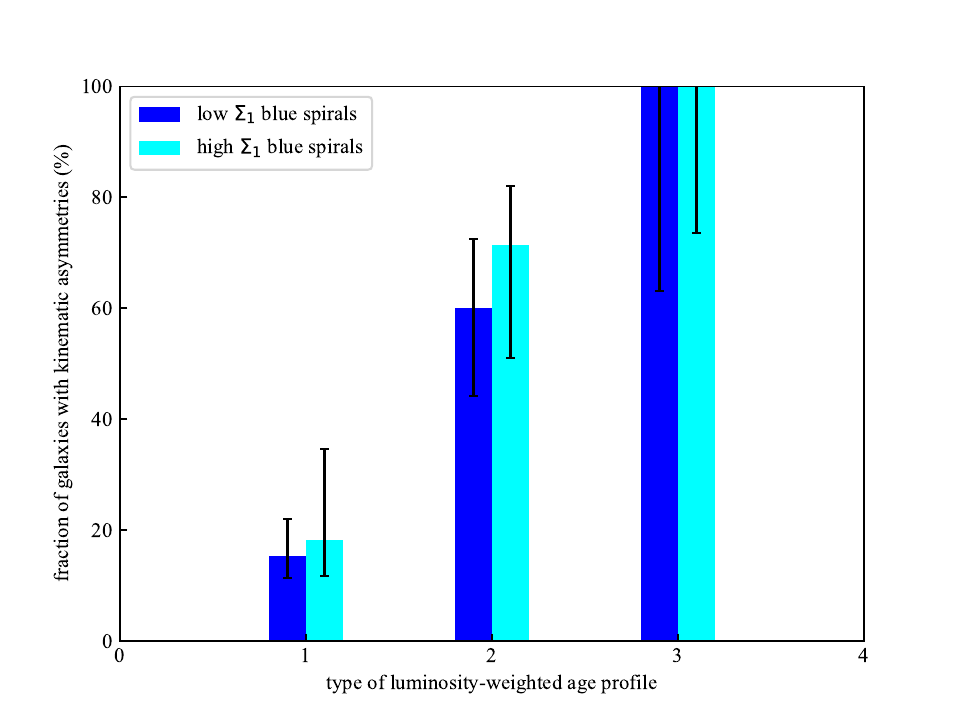}
	\caption{\label{KinAsymdis.eps} Fraction of galaxies with medium to
	high asymmetry in the H$\alpha$ velocity field, i.e., $0.027 <
	\bar{v}_{\rm asym} < 0.316$, as a function of the type of the
	luminosity-weighted age profiles for low $\Sigma_1$ (blue) and high
	$\Sigma_1$ (cyan) blue spirals. The x-axis represents the type of the
	age profiles, number coded by 1, 2 and 3.  One stands for a decreasing
	age profile, 2 stands for an increasing profile, and 3 stands for the
	age profile with a peak (i.e., ``turnover'' feature). The error bars
	represent the 1$\sigma$ binomial confidence limits based on the method
	of \citet{Cameron2011}.}
\end{figure}

The kinematic asymmetry in H$\alpha$ velocity field can probe the
perturbance suffered by a galaxy directly. We adopt the three intervals of the
kinematic asymmetry parameter used in \citet{Feng2020} to characterize our
samples of galaxies. Specifically, the three intervals of $0.007 < \bar{v}_{\rm
asym} < 0.027$, $0.027 < \bar{v}_{\rm asym} < 0.041$ and $0.041 < \bar{v}_{\rm
asym} < 0.316$ correspond to low, medium and high asymmetry in the H$\alpha$
velocity field, respectively. We then evaluate the association between
kinematic asymmetry and the luminosity-weighted age profile. Among the 12 high
$\Sigma_1$ blue spiral galaxies with medium to high asymmetry, except the 2
galaxies with the lowest asymmetry, the other 10 show either increasing age
profiles or ``turnover'' features in the age profiles. The other way around, we
also examine the fraction of galaxies with medium to high asymmetry as a
function of the age profile type, as shown in Figure~\ref{KinAsymdis.eps}. It
is obvious that the galaxies with a decreasing age profile have the least
fraction of galaxies with kinematic asymmetries, and more than 75\% (60\%) of
high (low) $\Sigma_1$ blue spirals with an increasing age profile or ``turnover''
feature show kinematic asymmetries. This suggests that the asymmetries are
closely linked with the age profiles for both types of blue spirals. By
examination, we find that barred galaxies show more kinematic asymmetry in
H$\alpha$ velocity field, consistent with the finding of \citet{Feng2022}. But
a large fraction of the kinematic asymmetries are also produced by non-barred
galaxies, some of which show morphological disturbances or are involved in pair
or group systems. In summary, optical morphologies and gas kinematics provide
consistent results that blue spirals with young centers show much more
asymmetries than those with old centers. Specifically, bars and galaxy
interactions that disturbed the gas velocity fields play an important role in
transporting gas inwards for the blue spirals with younger centers. The
effect of bars on gas inflow has been demonstrated in many studies \citep[and
references therein]{Yu2022}.  We notice that these previous studies mainly
proved the essential role of bars in the growth of pseudo bulges, i.e., the
secular evolution of disk galaxies.  In this work, we find that bars are also
able to drive gas inflow and trigger new star formation in massive disk
galaxies with dense cores.

In such a gas-inflow scenario, a depressed central gas-phase metallicity may be
expected, which are often seen in galaxy pairs or mergers
\citep[e.g.,][]{Kewley2006, Ellison2008, Rupke2008, Peeples2009, Scudder2012,
Ellison2013, Guo2016}. It has long been known that galaxy luminosity or stellar
mass is positively correlated with the gas-phase metallicity on a global scale
\citep[e.g.,][]{Lequeux1979,Tremonti2004}. In recent years, the development of
IFU observations has enabled the confirmation of a similar correlation between
stellar mass surface density and gas phase metallicity on local scales
\citep[e.g,][]{Rosales-Ortega2012,Sanchez2013,Barrera-Ballesteros2016}.

\begin{figure}
	\plottwo{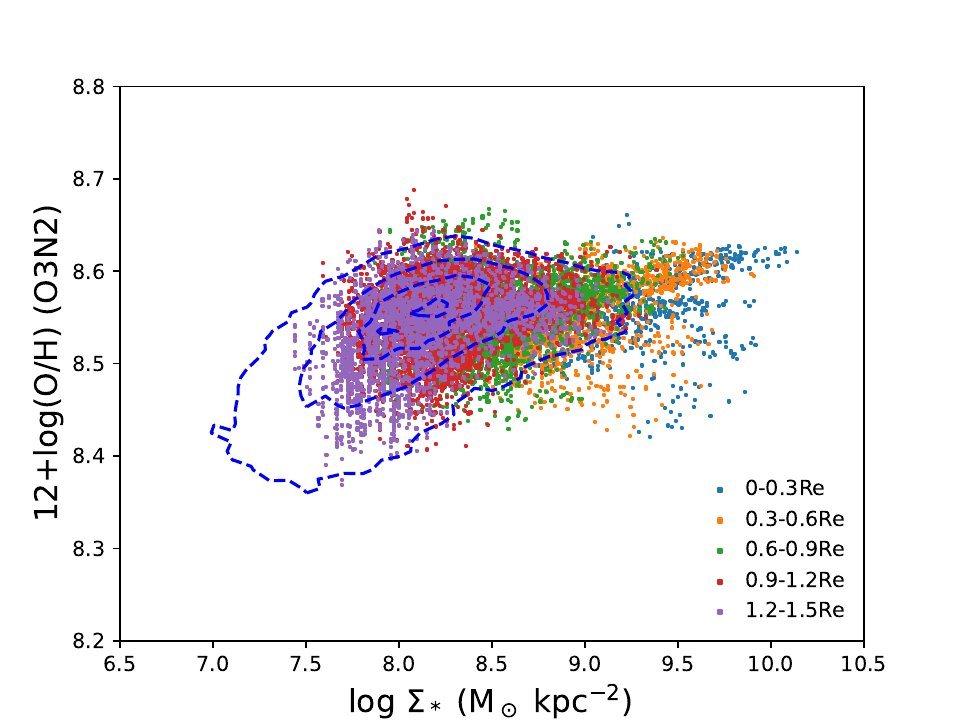}{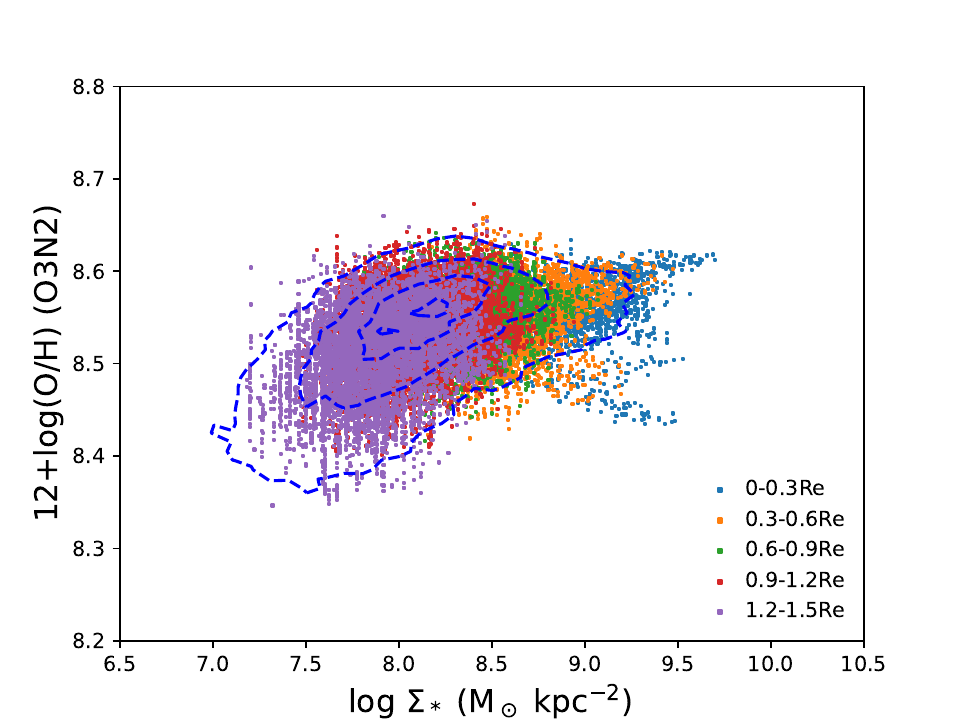}
	\caption{\label{spatiallyresolved_logoh.eps} Spatially-resolved stellar
	mass surface density versus gas-phase metallicity relations of
	star-forming spaxels for high ({\it left}) and low ({\it right})
	$\Sigma_1$ blue spirals with younger centers i.e., increasing or
	``turnover'' age profiles. The data points are color-coded according to radial
	distance to galactic center. The blue contours represent the number
	density distribution drawn from all the star-forming spaxels in low
	$\Sigma_1$ blue spirals. The contours enclose 5\%, 30\%, 68\% and 95\%
	of the sample, respectively.} 
\end{figure}

Following the widely used method, we search for possible metallicity dilution
in the blue spirals with younger centers, by studying the spatially-resolved
stellar mass surface density versus gas-phase metallicity relations in
Figure~\ref{spatiallyresolved_logoh.eps}. To obtain reliable oxygen abundance,
the most popular probe for the gas phase metallicity, only
[SII]/H$\alpha$-selected star-forming spaxels are used. The oxygen abundance
was calculated using the O3N2 calibration presented by the equation (2) in
\citet{Marino2013}. The data points in the left and right panels of
Figure~\ref{spatiallyresolved_logoh.eps} represent star-forming spaxels in high
$\Sigma_1$ and low $\Sigma_1$ blue spirals, respectively. Both panels show that
the data points in the central regions, represented by blue color, have the
highest mass surface densities, but the majority of them do not show
significantly lower gas-phase metallicities compared to the blue contours that
represent the distribution of the entire population of low $\Sigma_1$ blue
spirals. In fact, the gas phase metallicity is the result of the competition
between gas inflow/outflow, star formation and supernova feedback. It is only
when the metal-poor gas inflow is strong and the star formation lasts not long
enough to enrich the ISM via supernova explosion that the galaxy could show
central metallicity deficiency \citep{Montuori2010}. Such metallicity
deficiency is often seen in galaxy merging systems and evolves as the merger
proceeds. Actually, even in galaxy pairs, both metallicity dilution and
enrichment are found \citep{Omori2022}. In a study of CALIFA galaxies, the
authors found that the oxygen abundances are only slightly lower in tidally
perturbed galaxies than the control galaxies \citep{Morales-Vargas2021}. These
studies confirm the complexity of the physical processes imprinted on the gas
phase metallicity. In the case of gas inflow triggered by galactic bars, the
gas phase metallicity should be determined by similar physical processes, i.e.,
gas migration, star formation and the related supernova explosion etc, which
deserves a separate detailed study. Therefore, the normal oxygen abundance
shown in the central regions of high $\Sigma_1$ blue spirals does not provide
us with additional supportive evidence for gas inflow, but it does not conflict
with the gas-inflow scenario either.

On the other hand, we find that the properties of high $\Sigma_1$ blue spirals
with younger centers are also consistent with the fading counterparts of
Luminous Infrared Galaxies (LIRGs) with spiral morphologies (Guo et al., in
prep). In the parallel study, we find that spiral LIRGs also show dense (i.e.,
high $\Sigma_1$), younger and less $\alpha$-element enhanced centers but they host
much more intensive star formation. A more detailed comparison between high
$\Sigma_1$ blue spirals and spiral LIRGs will be performed in a forth-coming
paper.

For the remaining high $\Sigma_1$ blue spirals, which show old centers and low
sSFR in the center, we cannot distinguish their origins. Their properties are
seemingly consist with the inside-out growth scenario. Furthermore, they mostly
do not show kinematic asymmetries in H$\alpha$ velocity field, as illustrated
in Figure~\ref{KinAsymdis.eps}, although some of them have bars. However, we
could not rule out a rejuvenation origin if some minor disturbances caused by a
fly-by event trigger a small amount of gas inflow and star formation on a
galactic scale.

\section{SUMMARY and CONCLUSIONS}

The stellar mass -- $\Sigma_1$ relation followed by quiescent galaxies suggests
a close connection between galaxy quenching and the buildup of a dense core.
However, high $\Sigma_1$ blue spirals do not seem to fit the picture of galaxy
evolution drawn from this relation. They have assembled dense cores but are
still forming new stars. Based on the final data release of the SDSS-IV MaNGA IFU
data, we make an effort to shed light on the origins of massive ($ >
10^{10.5}{\rm M}_{\odot}$) high $\Sigma_1$ blue spirals at $0.02 < z < 0.05$,
by comparing the spatially-resolved stellar populations and star formation
properties between high $\Sigma_1$ and low $\Sigma_1$ blue spirals
with similar mass and redshift. The main results are summarized
below.

\begin{enumerate}
\item Both low $\Sigma_1$ and high $\Sigma_1$ blue spirals show large diversities
	in stellar population and star formation properties. The
		luminosity-weighted age profiles of blue spirals can be roughly
		divided into three categories, including a descending profile (i.e.,
		an older center and younger disk), an increasing profile (i.e.,
		a younger center and older disk) and a profile with a
		``turnover'' feature (a younger center and younger disk with an
		age peak in between). Compared to their low $\Sigma_1$
		counterparts, high $\Sigma_1$ blue spirals contain a larger
		fraction of galaxies with younger centers including galaxies
		with either an increasing profile or a profile with
		``turnover'' features ( $\sim 55\%$ versus $\sim 30\%$). These
	galaxies generally show smaller Mgb/${\rm \langle Fe \rangle}$ in the
centers and larger Mgb/${\rm \langle Fe \rangle}$ in the outer parts.  

\item Compared with their low $\Sigma_1$ counterparts, high $\Sigma_1$
	blue spirals possess a greater portion of galaxies with relatively
		large areal fractions of [NII]/H$\alpha$-identified composite
		regions. The composite regions prefer surrounding areas of bars
		or the outer disks of galaxies with rising age profiles. This
		suggests that the star formation in these galaxies with
		``younger'' centers is carrying on across the galaxy, but
		``pure star formation'' is more concentrated in the central
		regions.

\item The sSFR profiles show a good correspondence with the age profiles.
	Therefore, high $\Sigma_1$ blue spirals are composed of a larger
		fraction of galaxies with higher central sSFRs than low
		$\Sigma_1$ blue spirals ($\sim 70\%$ versus $\sim 40\%$).
		However, both the azimuthally averaged sSFR profiles and the
		spaxel-based star formation main sequence relations show that
		the star formation activities in the central regions of high
		$\Sigma_1$ blue spirals are not in a starburst mode.

\item Both the optical morphologies and the kinematic asymmetries in
	H$\alpha$ velocity field are closely linked with the shapes of the
		luminosity-weighted age profiles. More than 75\% of high
		$\Sigma_1$ blue spirals with increasing age profiles or
		“turnover” features show kinematic asymmetries. Bars or galaxy
		interactions should be responsible for the asymmetries shown in
		the H$\alpha$ velocity field.

\item High $\Sigma_1$ blue spirals that possess old centers and young disks do
	not show kinematic asymmetries, although some of them have bars.

\end{enumerate}

The properties summarized above suggest that more than half of the high $\Sigma_1$
blue spirals are likely to be rejuvenated systems.  Bars or galaxy interactions
drove the gas to flow into the galactic centers and triggered new star
formation mostly in the centers and also in the disks. However, we caution that
a direct identification of the rejuvenation from the stellar populations is
challenging. Although \citet{Zhang2022} proposed a method, its applicability to
real observational data still suffers from some difficulties. The remaining
high $\Sigma_1$ blue spirals have properties consistent with the inside-out
growth scenario. 

\begin{acknowledgments}

	We thank the anonymous referee for his/her constructive comments that
	significantly improved the paper.  This work is supported by the
	National Key Research and Development Program of China (No.
	2022YFA1602901) and the National Natural Science Foundation of China
	(NSFC, No. 12141301, 11733002). C.N.Hao also acknowledges the support
	from the science research grants from the China Manned Space Project
	with NO. CMS-CSST-2021-A04, CMS-CSST-2021-A07 and CMS-CSST-2021-B02.
	Funding for the Sloan Digital Sky Survey IV has been provided by the
	Alfred P. Sloan Foundation, the U.S.  Department of Energy Office of
	Science, and the Participating Institutions. SDSS-IV acknowledges
	support and resources from the Center for High-Performance Computing at
	the University of Utah. The SDSS web site is www.sdss.org.  SDSS-IV is
	managed by the Astrophysical Research Consortium for the Participating
	Institutions of the SDSS Collaboration including the Brazilian
	Participation Group, the Carnegie Institution for Science, Carnegie
	Mellon University, the Chilean Participation Group, the French
	Participation Group, Harvard-Smithsonian Center for Astrophysics,
	Instituto de Astrof\'isica de Canarias, The Johns Hopkins University,
	Kavli Institute for the Physics and Mathematics of the Universe (IPMU)
	/ University of Tokyo, the Korean Participation Group, Lawrence
	Berkeley National Laboratory, Leibniz Institut f\"ur Astrophysik
	Potsdam (AIP), Max-Planck-Institut f\"ur Astronomie (MPIA Heidelberg),
	Max-Planck-Institut f\"ur Astrophysik (MPA Garching),
	Max-Planck-Institut f\"ur Extraterrestrische Physik (MPE), National
	Astronomical Observatories of China, New Mexico State University, New
	York University, University of Notre Dame, Observat\'ario Nacional /
	MCTI, The Ohio State University, Pennsylvania State University,
	Shanghai Astronomical Observatory, United Kingdom Participation Group,
	Universidad Nacional Aut\'onoma de M\'exico, University of Arizona,
	University of Colorado Boulder, University of Oxford, University of
	Portsmouth, University of Utah, University of Virginia, University of
	Washington, University of Wisconsin, Vanderbilt University, and Yale
	University.  This project makes use of the MaNGA-Pipe3D data products.
	We thank the IA-UNAM MaNGA team for creating this catalogue, and the
	ConaCyt-180125 project for supporting them.  \end{acknowledgments}

\bibliography{highsigma1bib}{}

\begin{thebibliography}{}
\expandafter\ifx\csname natexlab\endcsname\relax\def\natexlab#1{#1}\fi
\providecommand{\url}[1]{\href{#1}{#1}}
\providecommand{\dodoi}[1]{doi:~\href{http://doi.org/#1}{\nolinkurl{#1}}}
\providecommand{\doeprint}[1]{\href{http://ascl.net/#1}{\nolinkurl{http://ascl.net/#1}}}
\providecommand{\doarXiv}[1]{\href{https://arxiv.org/abs/#1}{\nolinkurl{https://arxiv.org/abs/#1}}}

\bibitem[{{Abdurro'uf} {et~al.}(2022){Abdurro'uf}, {Accetta}, {Aerts}, {Silva
  Aguirre}, {Ahumada}, {Ajgaonkar}, {Filiz Ak}, {Alam}, {Allende Prieto},
  {Almeida}, {Anders}, {Anderson}, {Andrews}, {Anguiano}, {Aquino-Ort{\'\i}z},
  {Arag{\'o}n-Salamanca}, {Argudo-Fern{\'a}ndez}, {Ata}, {Aubert},
  {Avila-Reese}, {Badenes}, {Barb{\'a}}, {Barger}, {Barrera-Ballesteros},
  {Beaton}, {Beers}, {Belfiore}, {Bender}, {Bernardi}, {Bershady}, {Beutler},
  {Bidin}, {Bird}, {Bizyaev}, {Blanc}, {Blanton}, {Boardman}, {Bolton},
  {Boquien}, {Borissova}, {Bovy}, {Brandt}, {Brown}, {Brownstein}, {Brusa},
  {Buchner}, {Bundy}, {Burchett}, {Bureau}, {Burgasser}, {Cabang}, {Campbell},
  {Cappellari}, {Carlberg}, {Wanderley}, {Carrera}, {Cash}, {Chen}, {Chen},
  {Cherinka}, {Chiappini}, {Choi}, {Chojnowski}, {Chung}, {Clerc}, {Cohen},
  {Comerford}, {Comparat}, {da Costa}, {Covey}, {Crane}, {Cruz-Gonzalez},
  {Culhane}, {Cunha}, {Dai}, {Damke}, {Darling}, {Davidson}, {Davies},
  {Dawson}, {De Lee}, {Diamond-Stanic}, {Cano-D{\'\i}az}, {S{\'a}nchez},
  {Donor}, {Duckworth}, {Dwelly}, {Eisenstein}, {Elsworth}, {Emsellem},
  {Eracleous}, {Escoffier}, {Fan}, {Farr}, {Feng}, {Fern{\'a}ndez-Trincado},
  {Feuillet}, {Filipp}, {Fillingham}, {Frinchaboy}, {Fromenteau}, {Galbany},
  {Garc{\'\i}a}, {Garc{\'\i}a-Hern{\'a}ndez}, {Ge}, {Geisler}, {Gelfand},
  {G{\'e}ron}, {Gibson}, {Goddy}, {Godoy-Rivera}, {Grabowski}, {Green},
  {Greener}, {Grier}, {Griffith}, {Guo}, {Guy}, {Hadjara}, {Harding},
  {Hasselquist}, {Hayes}, {Hearty}, {Hern{\'a}ndez}, {Hill}, {Hogg},
  {Holtzman}, {Horta}, {Hsieh}, {Hsu}, {Hsu}, {Huber}, {Huertas-Company},
  {Hutchinson}, {Hwang}, {Ibarra-Medel}, {Chitham}, {Ilha}, {Imig}, {Jaekle},
  {Jayasinghe}, {Ji}, {Johnson}, {Jones}, {J{\"o}nsson}, {Katkov}, {Khalatyan},
  {Kinemuchi}, {Kisku}, {Knapen}, {Kneib}, {Kollmeier}, {Kong}, {Kounkel},
  {Kreckel}, {Krishnarao}, {Lacerna}, {Lane}, {Langgin}, {Lavender}, {Law},
  {Lazarz}, {Leung}, {Leung}, {Lewis}, {Li}, {Li}, {Lian}, {Liang}, {Lin},
  {Lin}, {Lin}, {Lintott}, {Long}, {Longa-Pe{\~n}a}, {L{\'o}pez-Cob{\'a}},
  {Lu}, {Lundgren}, {Luo}, {Mackereth}, {de la Macorra}, {Mahadevan},
  {Majewski}, {Manchado}, {Mandeville}, {Maraston}, {Margalef-Bentabol},
  {Masseron}, {Masters}, {Mathur}, {McDermid}, {Mckay}, {Merloni},
  {Merrifield}, {Meszaros}, {Miglio}, {Di Mille}, {Minniti}, {Minsley},
  {Monachesi}, {Moon}, {Mosser}, {Mulchaey}, {Muna}, {Mu{\~n}oz}, {Myers},
  {Myers}, {Nadathur}, {Nair}, {Nandra}, {Neumann}, {Newman}, {Nidever},
  {Nikakhtar}, {Nitschelm}, {O'Connell}, {Garma-Oehmichen}, {Luan Souza de
  Oliveira}, {Olney}, {Oravetz}, {Ortigoza-Urdaneta}, {Osorio}, {Otter},
  {Pace}, {Padilla}, {Pan}, {Pan}, {Parikh}, {Parker}, {Peirani}, {Pe{\~n}a
  Ram{\'\i}rez}, {Penny}, {Percival}, {Perez-Fournon}, {Pinsonneault},
  {Poidevin}, {Poovelil}, {Price-Whelan}, {B{\'a}rbara de Andrade Queiroz},
  {Raddick}, {Ray}, {Rembold}, {Riddle}, {Riffel}, {Riffel}, {Rix}, {Robin},
  {Rodr{\'\i}guez-Puebla}, {Roman-Lopes}, {Rom{\'a}n-Z{\'u}{\~n}iga}, {Rose},
  {Ross}, {Rossi}, {Rubin}, {Salvato}, {S{\'a}nchez}, {S{\'a}nchez-Gallego},
  {Sanderson}, {Santana Rojas}, {Sarceno}, {Sarmiento}, {Sayres}, {Sazonova},
  {Schaefer}, {Schiavon}, {Schlegel}, {Schneider}, {Schultheis}, {Schwope},
  {Serenelli}, {Serna}, {Shao}, {Shapiro}, {Sharma}, {Shen}, {Shetrone}, {Shu},
  {Simon}, {Skrutskie}, {Smethurst}, {Smith}, {Sobeck}, {Spoo}, {Sprague},
  {Stark}, {Stassun}, {Steinmetz}, {Stello}, {Stone-Martinez},
  {Storchi-Bergmann}, {Stringfellow}, {Stutz}, {Su}, {Taghizadeh-Popp},
  {Talbot}, {Tayar}, {Telles}, {Teske}, {Thakar}, {Theissen}, {Tkachenko},
  {Thomas}, {Tojeiro}, {Hernandez Toledo}, {Troup}, {Trump}, {Trussler},
  {Turner}, {Tuttle}, {Unda-Sanzana}, {V{\'a}zquez-Mata}, {Valentini},
  {Valenzuela}, {Vargas-Gonz{\'a}lez}, {Vargas-Maga{\~n}a}, {Alfaro},
  {Villanova}, {Vincenzo}, {Wake}, {Warfield}, {Washington}, {Weaver},
  {Weijmans}, {Weinberg}, {Weiss}, {Westfall}, {Wild}, {Wilde}, {Wilson},
  {Wilson}, {Wilson}, {Wolf}, {Wood-Vasey}, {Yan}, {Zamora}, {Zasowski},
  {Zhang}, {Zhao}, {Zheng}, {Zheng}, \& {Zhu}}]{Abdurrouf2022}
{Abdurro'uf}, {Accetta}, K., {Aerts}, C., {et~al.} 2022, \apjs, 259, 35,
  \dodoi{10.3847/1538-4365/ac4414}

\bibitem[{{Baldwin} {et~al.}(1981){Baldwin}, {Phillips}, \&
  {Terlevich}}]{Baldwin1981}
{Baldwin}, J.~A., {Phillips}, M.~M., \& {Terlevich}, R. 1981, \pasp, 93, 5,
  \dodoi{10.1086/130766}

\bibitem[{{Barrera-Ballesteros} {et~al.}(2016){Barrera-Ballesteros}, {Heckman},
  {Zhu}, {Zakamska}, {S{\'a}nchez}, {Law}, {Wake}, {Green}, {Bizyaev},
  {Oravetz}, {Simmons}, {Malanushenko}, {Pan}, {Roman Lopes}, \&
  {Lane}}]{Barrera-Ballesteros2016}
{Barrera-Ballesteros}, J.~K., {Heckman}, T.~M., {Zhu}, G.~B., {et~al.} 2016,
  \mnras, 463, 2513, \dodoi{10.1093/mnras/stw1984}

\bibitem[{{Barro} {et~al.}(2017){Barro}, {Faber}, {Koo}, {Dekel}, {Fang},
  {Trump}, {P{\'e}rez-Gonz{\'a}lez}, {Pacifici}, {Primack}, {Somerville},
  {Yan}, {Guo}, {Liu}, {Ceverino}, {Kocevski}, \& {McGrath}}]{Barro2017}
{Barro}, G., {Faber}, S.~M., {Koo}, D.~C., {et~al.} 2017, \apj, 840, 47,
  \dodoi{10.3847/1538-4357/aa6b05}

\bibitem[{{Belfiore} {et~al.}(2019){Belfiore}, {Westfall}, {Schaefer},
  {Cappellari}, {Ji}, {Bershady}, {Tremonti}, {Law}, {Yan}, {Bundy}, {Shetty},
  {Drory}, {Thomas}, {Emsellem}, \& {S{\'a}nchez}}]{Belfiore2019}
{Belfiore}, F., {Westfall}, K.~B., {Schaefer}, A., {et~al.} 2019, \aj, 158,
  160, \dodoi{10.3847/1538-3881/ab3e4e}

\bibitem[{{Blanton} {et~al.}(2011){Blanton}, {Kazin}, {Muna}, {Weaver}, \&
  {Price-Whelan}}]{Blanton2011}
{Blanton}, M.~R., {Kazin}, E., {Muna}, D., {Weaver}, B.~A., \& {Price-Whelan},
  A. 2011, \aj, 142, 31, \dodoi{10.1088/0004-6256/142/1/31}

\bibitem[{{Bundy} {et~al.}(2015){Bundy}, {Bershady}, {Law}, {Yan}, {Drory},
  {MacDonald}, {Wake}, {Cherinka}, {S{\'a}nchez-Gallego}, {Weijmans}, {Thomas},
  {Tremonti}, {Masters}, {Coccato}, {Diamond-Stanic}, {Arag{\'o}n-Salamanca},
  {Avila-Reese}, {Badenes}, {Falc{\'o}n-Barroso}, {Belfiore}, {Bizyaev},
  {Blanc}, {Bland-Hawthorn}, {Blanton}, {Brownstein}, {Byler}, {Cappellari},
  {Conroy}, {Dutton}, {Emsellem}, {Etherington}, {Frinchaboy}, {Fu}, {Gunn},
  {Harding}, {Johnston}, {Kauffmann}, {Kinemuchi}, {Klaene}, {Knapen},
  {Leauthaud}, {Li}, {Lin}, {Maiolino}, {Malanushenko}, {Malanushenko}, {Mao},
  {Maraston}, {McDermid}, {Merrifield}, {Nichol}, {Oravetz}, {Pan}, {Parejko},
  {Sanchez}, {Schlegel}, {Simmons}, {Steele}, {Steinmetz}, {Thanjavur},
  {Thompson}, {Tinker}, {van den Bosch}, {Westfall}, {Wilkinson}, {Wright},
  {Xiao}, \& {Zhang}}]{Bundy2015}
{Bundy}, K., {Bershady}, M.~A., {Law}, D.~R., {et~al.} 2015, \apj, 798, 7,
  \dodoi{10.1088/0004-637X/798/1/7}

\bibitem[{{Cameron}(2011)}]{Cameron2011}
{Cameron}, E. 2011, \pasa, 28, 128, \dodoi{10.1071/AS10046}

\bibitem[{{Chabrier}(2003)}]{Chabrier2003}
{Chabrier}, G. 2003, \pasp, 115, 763, \dodoi{10.1086/376392}

\bibitem[{{Cheung} {et~al.}(2012){Cheung}, {Faber}, {Koo}, {Dutton}, {Simard},
  {McGrath}, {Huang}, {Bell}, {Dekel}, {Fang}, {Salim}, {Barro}, {Bundy},
  {Coil}, {Cooper}, {Conselice}, {Davis}, {Dom{\'\i}nguez}, {Kassin},
  {Kocevski}, {Koekemoer}, {Lin}, {Lotz}, {Newman}, {Phillips}, {Rosario},
  {Weiner}, \& {Willmer}}]{Cheung2012}
{Cheung}, E., {Faber}, S.~M., {Koo}, D.~C., {et~al.} 2012, \apj, 760, 131,
  \dodoi{10.1088/0004-637X/760/2/131}

\bibitem[{{Dekel} {et~al.}(2019){Dekel}, {Lapiner}, \& {Dubois}}]{Dekel2019}
{Dekel}, A., {Lapiner}, S., \& {Dubois}, Y. 2019, arXiv e-prints,
  arXiv:1904.08431.
\newblock \doarXiv{1904.08431}

\bibitem[{{Dey} {et~al.}(2019){Dey}, {Schlegel}, {Lang}, {Blum}, {Burleigh},
  {Fan}, {Findlay}, {Finkbeiner}, {Herrera}, {Juneau}, {Landriau}, {Levi},
  {McGreer}, {Meisner}, {Myers}, {Moustakas}, {Nugent}, {Patej}, {Schlafly},
  {Walker}, {Valdes}, {Weaver}, {Y{\`e}che}, {Zou}, {Zhou}, {Abareshi},
  {Abbott}, {Abolfathi}, {Aguilera}, {Alam}, {Allen}, {Alvarez}, {Annis},
  {Ansarinejad}, {Aubert}, {Beechert}, {Bell}, {BenZvi}, {Beutler}, {Bielby},
  {Bolton}, {Brice{\~n}o}, {Buckley-Geer}, {Butler}, {Calamida}, {Carlberg},
  {Carter}, {Casas}, {Castander}, {Choi}, {Comparat}, {Cukanovaite}, {Delubac},
  {DeVries}, {Dey}, {Dhungana}, {Dickinson}, {Ding}, {Donaldson}, {Duan},
  {Duckworth}, {Eftekharzadeh}, {Eisenstein}, {Etourneau}, {Fagrelius},
  {Farihi}, {Fitzpatrick}, {Font-Ribera}, {Fulmer}, {G{\"a}nsicke},
  {Gaztanaga}, {George}, {Gerdes}, {Gontcho}, {Gorgoni}, {Green}, {Guy},
  {Harmer}, {Hernandez}, {Honscheid}, {Huang}, {James}, {Jannuzi}, {Jiang},
  {Joyce}, {Karcher}, {Karkar}, {Kehoe}, {Kneib}, {Kueter-Young}, {Lan},
  {Lauer}, {Le Guillou}, {Le Van Suu}, {Lee}, {Lesser}, {Perreault Levasseur},
  {Li}, {Mann}, {Marshall}, {Mart{\'\i}nez-V{\'a}zquez}, {Martini}, {du Mas des
  Bourboux}, {McManus}, {Meier}, {M{\'e}nard}, {Metcalfe},
  {Mu{\~n}oz-Guti{\'e}rrez}, {Najita}, {Napier}, {Narayan}, {Newman}, {Nie},
  {Nord}, {Norman}, {Olsen}, {Paat}, {Palanque-Delabrouille}, {Peng},
  {Poppett}, {Poremba}, {Prakash}, {Rabinowitz}, {Raichoor}, {Rezaie},
  {Robertson}, {Roe}, {Ross}, {Ross}, {Rudnick}, {Safonova}, {Saha},
  {S{\'a}nchez}, {Savary}, {Schweiker}, {Scott}, {Seo}, {Shan}, {Silva},
  {Slepian}, {Soto}, {Sprayberry}, {Staten}, {Stillman}, {Stupak}, {Summers},
  {Sien Tie}, {Tirado}, {Vargas-Maga{\~n}a}, {Vivas}, {Wechsler}, {Williams},
  {Yang}, {Yang}, {Yapici}, {Zaritsky}, {Zenteno}, {Zhang}, {Zhang}, {Zhou}, \&
  {Zhou}}]{Dey2019}
{Dey}, A., {Schlegel}, D.~J., {Lang}, D., {et~al.} 2019, \aj, 157, 168,
  \dodoi{10.3847/1538-3881/ab089d}

\bibitem[{{Ellison} {et~al.}(2013){Ellison}, {Mendel}, {Patton}, \&
  {Scudder}}]{Ellison2013}
{Ellison}, S.~L., {Mendel}, J.~T., {Patton}, D.~R., \& {Scudder}, J.~M. 2013,
  \mnras, 435, 3627, \dodoi{10.1093/mnras/stt1562}

\bibitem[{{Ellison} {et~al.}(2008){Ellison}, {Patton}, {Simard}, \&
  {McConnachie}}]{Ellison2008}
{Ellison}, S.~L., {Patton}, D.~R., {Simard}, L., \& {McConnachie}, A.~W. 2008,
  \aj, 135, 1877, \dodoi{10.1088/0004-6256/135/5/1877}

\bibitem[{{Fang} {et~al.}(2013){Fang}, {Faber}, {Koo}, \& {Dekel}}]{Fang2013}
{Fang}, J.~J., {Faber}, S.~M., {Koo}, D.~C., \& {Dekel}, A. 2013, \apj, 776,
  63, \dodoi{10.1088/0004-637X/776/1/63}

\bibitem[{{Feng} {et~al.}(2022){Feng}, {Shen}, {Yuan}, {Dai}, \&
  {Masters}}]{Feng2022}
{Feng}, S., {Shen}, S.-Y., {Yuan}, F.-T., {Dai}, Y.~S., \& {Masters}, K.~L.
  2022, \apjs, 262, 6, \dodoi{10.3847/1538-4365/ac80f2}

\bibitem[{{Feng} {et~al.}(2020){Feng}, {Shen}, {Yuan}, {Riffel}, \&
  {Pan}}]{Feng2020}
{Feng}, S., {Shen}, S.-Y., {Yuan}, F.-T., {Riffel}, R.~A., \& {Pan}, K. 2020,
  \apjl, 892, L20, \dodoi{10.3847/2041-8213/ab7dba}

\bibitem[{{Guo} {et~al.}(2020){Guo}, {Hao}, {Xia}, {Shi}, {Chen}, {Li}, \&
  {Gu}}]{Guo2020}
{Guo}, R., {Hao}, C.-N., {Xia}, X., {et~al.} 2020, \apj, 897, 162,
  \dodoi{10.3847/1538-4357/ab9b75}

\bibitem[{{Guo} {et~al.}(2016){Guo}, {Hao}, {Xia}, {Wei}, \& {Guo}}]{Guo2016}
{Guo}, R., {Hao}, C.-N., {Xia}, X.-Y., {Wei}, P., \& {Guo}, X. 2016, Research
  in Astronomy and Astrophysics, 16, 113, \dodoi{10.1088/1674-4527/16/7/113}

\bibitem[{{Hao} {et~al.}(2019){Hao}, {Shi}, {Chen}, {Xia}, {Gu}, {Guo}, {Yu},
  \& {Li}}]{Hao2019}
{Hao}, C.-N., {Shi}, Y., {Chen}, Y., {et~al.} 2019, \apjl, 883, L36,
  \dodoi{10.3847/2041-8213/ab42e5}

\bibitem[{{Hart} {et~al.}(2016){Hart}, {Bamford}, {Willett}, {Masters},
  {Cardamone}, {Lintott}, {Mackay}, {Nichol}, {Rosslowe}, {Simmons}, \&
  {Smethurst}}]{Hart2016}
{Hart}, R.~E., {Bamford}, S.~P., {Willett}, K.~W., {et~al.} 2016, \mnras, 461,
  3663, \dodoi{10.1093/mnras/stw1588}

\bibitem[{{Kauffmann} {et~al.}(2003){Kauffmann}, {Heckman}, {Tremonti},
  {Brinchmann}, {Charlot}, {White}, {Ridgway}, {Brinkmann}, {Fukugita}, {Hall},
  {Ivezi{\'c}}, {Richards}, \& {Schneider}}]{Kauffmann2003}
{Kauffmann}, G., {Heckman}, T.~M., {Tremonti}, C., {et~al.} 2003, \mnras, 346,
  1055, \dodoi{10.1111/j.1365-2966.2003.07154.x}

\bibitem[{{Kennicutt} {et~al.}(2009){Kennicutt}, {Hao}, {Calzetti},
  {Moustakas}, {Dale}, {Bendo}, {Engelbracht}, {Johnson}, \&
  {Lee}}]{Kennicutt2009}
{Kennicutt}, Robert~C., J., {Hao}, C.-N., {Calzetti}, D., {et~al.} 2009, \apj,
  703, 1672, \dodoi{10.1088/0004-637X/703/2/1672}

\bibitem[{{Kewley} {et~al.}(2001){Kewley}, {Dopita}, {Sutherland}, {Heisler},
  \& {Trevena}}]{Kewley2001}
{Kewley}, L.~J., {Dopita}, M.~A., {Sutherland}, R.~S., {Heisler}, C.~A., \&
  {Trevena}, J. 2001, \apj, 556, 121, \dodoi{10.1086/321545}

\bibitem[{{Kewley} {et~al.}(2006){Kewley}, {Geller}, \& {Barton}}]{Kewley2006}
{Kewley}, L.~J., {Geller}, M.~J., \& {Barton}, E.~J. 2006, \aj, 131, 2004,
  \dodoi{10.1086/500295}

\bibitem[{{Krajnovi{\'c}} {et~al.}(2006){Krajnovi{\'c}}, {Cappellari}, {de
  Zeeuw}, \& {Copin}}]{Krajnovic2006}
{Krajnovi{\'c}}, D., {Cappellari}, M., {de Zeeuw}, P.~T., \& {Copin}, Y. 2006,
  \mnras, 366, 787, \dodoi{10.1111/j.1365-2966.2005.09902.x}

\bibitem[{{Lacerda} {et~al.}(2022){Lacerda}, {S{\'a}nchez},
  {Mej{\'\i}a-Narv{\'a}ez}, {Camps-Fari{\~n}a}, {Espinosa-Ponce},
  {Barrera-Ballesteros}, {Ibarra-Medel}, \& {Lugo-Aranda}}]{Lacerda2022}
{Lacerda}, E. A.~D., {S{\'a}nchez}, S.~F., {Mej{\'\i}a-Narv{\'a}ez}, A.,
  {et~al.} 2022, \na, 97, 101895, \dodoi{10.1016/j.newast.2022.101895}

\bibitem[{{Leitherer} {et~al.}(2014){Leitherer}, {Ekstr{\"o}m}, {Meynet},
  {Schaerer}, {Agienko}, \& {Levesque}}]{Leitherer2014}
{Leitherer}, C., {Ekstr{\"o}m}, S., {Meynet}, G., {et~al.} 2014, \apjs, 212,
  14, \dodoi{10.1088/0067-0049/212/1/14}

\bibitem[{{Leitherer} {et~al.}(2010){Leitherer}, {Ortiz Ot{\'a}lvaro},
  {Bresolin}, {Kudritzki}, {Lo Faro}, {Pauldrach}, {Pettini}, \&
  {Rix}}]{Leitherer2010}
{Leitherer}, C., {Ortiz Ot{\'a}lvaro}, P.~A., {Bresolin}, F., {et~al.} 2010,
  \apjs, 189, 309, \dodoi{10.1088/0067-0049/189/2/309}

\bibitem[{{Leitherer} {et~al.}(1999){Leitherer}, {Schaerer}, {Goldader},
  {Delgado}, {Robert}, {Kune}, {de Mello}, {Devost}, \&
  {Heckman}}]{Leitherer1999}
{Leitherer}, C., {Schaerer}, D., {Goldader}, J.~D., {et~al.} 1999, \apjs, 123,
  3, \dodoi{10.1086/313233}

\bibitem[{{Lequeux} {et~al.}(1979){Lequeux}, {Peimbert}, {Rayo}, {Serrano}, \&
  {Torres-Peimbert}}]{Lequeux1979}
{Lequeux}, J., {Peimbert}, M., {Rayo}, J.~F., {Serrano}, A., \&
  {Torres-Peimbert}, S. 1979, \aap, 80, 155

\bibitem[{{Lin} {et~al.}(2017){Lin}, {Li}, {He}, {Xiao}, \&
  {Wang}}]{LinLin2017}
{Lin}, L., {Li}, C., {He}, Y., {Xiao}, T., \& {Wang}, E. 2017, \apj, 838, 105,
  \dodoi{10.3847/1538-4357/aa657a}

\bibitem[{{Lin} {et~al.}(2020){Lin}, {Li}, {Du}, {Wang}, {Xiao}, {Bureau},
  {Fraser-McKelvie}, {Masters}, {Lin}, {Wake}, \& {Hao}}]{LinLin2020}
{Lin}, L., {Li}, C., {Du}, C., {et~al.} 2020, \mnras, 499, 1406,
  \dodoi{10.1093/mnras/staa2913}

\bibitem[{{Lintott} {et~al.}(2011){Lintott}, {Schawinski}, {Bamford}, {Slosar},
  {Land}, {Thomas}, {Edmondson}, {Masters}, {Nichol}, {Raddick}, {Szalay},
  {Andreescu}, {Murray}, \& {Vandenberg}}]{Lintott2011}
{Lintott}, C., {Schawinski}, K., {Bamford}, S., {et~al.} 2011, \mnras, 410,
  166, \dodoi{10.1111/j.1365-2966.2010.17432.x}

\bibitem[{{Lintott} {et~al.}(2008){Lintott}, {Schawinski}, {Slosar}, {Land},
  {Bamford}, {Thomas}, {Raddick}, {Nichol}, {Szalay}, {Andreescu}, {Murray}, \&
  {Vandenberg}}]{Lintott2008}
{Lintott}, C.~J., {Schawinski}, K., {Slosar}, A., {et~al.} 2008, \mnras, 389,
  1179, \dodoi{10.1111/j.1365-2966.2008.13689.x}

\bibitem[{{Marino} {et~al.}(2013){Marino}, {Rosales-Ortega}, {S{\'a}nchez},
  {Gil de Paz}, {V{\'\i}lchez}, {Miralles-Caballero}, {Kehrig},
  {P{\'e}rez-Montero}, {Stanishev}, {Iglesias-P{\'a}ramo}, {D{\'\i}az},
  {Castillo-Morales}, {Kennicutt}, {L{\'o}pez-S{\'a}nchez}, {Galbany},
  {Garc{\'\i}a-Benito}, {Mast}, {Mendez-Abreu}, {Monreal-Ibero}, {Husemann},
  {Walcher}, {Garc{\'\i}a-Lorenzo}, {Masegosa}, {Del Olmo Orozco},
  {Mour{\~a}o}, {Ziegler}, {Moll{\'a}}, {Papaderos},
  {S{\'a}nchez-Bl{\'a}zquez}, {Gonz{\'a}lez Delgado}, {Falc{\'o}n-Barroso},
  {Roth}, {van de Ven}, \& {Califa Team}}]{Marino2013}
{Marino}, R.~A., {Rosales-Ortega}, F.~F., {S{\'a}nchez}, S.~F., {et~al.} 2013,
  \aap, 559, A114, \dodoi{10.1051/0004-6361/201321956}

\bibitem[{{Mendel} {et~al.}(2014){Mendel}, {Simard}, {Palmer}, {Ellison}, \&
  {Patton}}]{Mendel2014}
{Mendel}, J.~T., {Simard}, L., {Palmer}, M., {Ellison}, S.~L., \& {Patton},
  D.~R. 2014, \apjs, 210, 3, \dodoi{10.1088/0067-0049/210/1/3}

\bibitem[{{Montuori} {et~al.}(2010){Montuori}, {Di Matteo}, {Lehnert},
  {Combes}, \& {Semelin}}]{Montuori2010}
{Montuori}, M., {Di Matteo}, P., {Lehnert}, M.~D., {Combes}, F., \& {Semelin},
  B. 2010, \aap, 518, A56, \dodoi{10.1051/0004-6361/201014304}

\bibitem[{{Morales-Vargas} {et~al.}(2021){Morales-Vargas}, {Torres-Papaqui},
  {Rosales-Ortega}, {Chow-Mart{\'\i}nez}, {Trejo-Alonso}, {Ortega-Minakata},
  {Robleto-Or{\'u}s}, {Romero-Cruz}, {Neri-Larios}, {Neri-Larios}, \& {Califa
  Survey Collaboration}}]{Morales-Vargas2021}
{Morales-Vargas}, A., {Torres-Papaqui}, J.~P., {Rosales-Ortega}, F.~F.,
  {et~al.} 2021, \mnras, 508, 4216, \dodoi{10.1093/mnras/stab2698}

\bibitem[{{O'Donnell}(1994)}]{Donnell1994}
{O'Donnell}, J.~E. 1994, \apj, 422, 158, \dodoi{10.1086/173713}

\bibitem[{{Omori} \& {Takeuchi}(2022)}]{Omori2022}
{Omori}, K.~C., \& {Takeuchi}, T.~T. 2022, \aap, 660, A145,
  \dodoi{10.1051/0004-6361/202142858}

\bibitem[{{Peeples} {et~al.}(2009){Peeples}, {Pogge}, \&
  {Stanek}}]{Peeples2009}
{Peeples}, M.~S., {Pogge}, R.~W., \& {Stanek}, K.~Z. 2009, \apj, 695, 259,
  \dodoi{10.1088/0004-637X/695/1/259}

\bibitem[{{Rosales-Ortega} {et~al.}(2012){Rosales-Ortega}, {S{\'a}nchez},
  {Iglesias-P{\'a}ramo}, {D{\'\i}az}, {V{\'\i}lchez}, {Bland-Hawthorn},
  {Husemann}, \& {Mast}}]{Rosales-Ortega2012}
{Rosales-Ortega}, F.~F., {S{\'a}nchez}, S.~F., {Iglesias-P{\'a}ramo}, J.,
  {et~al.} 2012, \apjl, 756, L31, \dodoi{10.1088/2041-8205/756/2/L31}

\bibitem[{{Rupke} {et~al.}(2008){Rupke}, {Veilleux}, \& {Baker}}]{Rupke2008}
{Rupke}, D. S.~N., {Veilleux}, S., \& {Baker}, A.~J. 2008, \apj, 674, 172,
  \dodoi{10.1086/522363}

\bibitem[{{Salim} {et~al.}(2018){Salim}, {Boquien}, \& {Lee}}]{Salim2018}
{Salim}, S., {Boquien}, M., \& {Lee}, J.~C. 2018, \apj, 859, 11,
  \dodoi{10.3847/1538-4357/aabf3c}

\bibitem[{{Salpeter}(1955)}]{Salpeter1955}
{Salpeter}, E.~E. 1955, \apj, 121, 161, \dodoi{10.1086/145971}

\bibitem[{{S{\'a}nchez} {et~al.}(2013){S{\'a}nchez}, {Rosales-Ortega},
  {Jungwiert}, {Iglesias-P{\'a}ramo}, {V{\'\i}lchez}, {Marino}, {Walcher},
  {Husemann}, {Mast}, {Monreal-Ibero}, {Cid Fernandes}, {P{\'e}rez},
  {Gonz{\'a}lez Delgado}, {Garc{\'\i}a-Benito}, {Galbany}, {van de Ven},
  {Jahnke}, {Flores}, {Bland-Hawthorn}, {L{\'o}pez-S{\'a}nchez}, {Stanishev},
  {Miralles-Caballero}, {D{\'\i}az}, {S{\'a}nchez-Blazquez}, {Moll{\'a}},
  {Gallazzi}, {Papaderos}, {Gomes}, {Gruel}, {P{\'e}rez}, {Ruiz-Lara},
  {Florido}, {de Lorenzo-C{\'a}ceres}, {Mendez-Abreu}, {Kehrig}, {Roth},
  {Ziegler}, {Alves}, {Wisotzki}, {Kupko}, {Quirrenbach}, {Bomans}, \& {Califa
  Collaboration}}]{Sanchez2013}
{S{\'a}nchez}, S.~F., {Rosales-Ortega}, F.~F., {Jungwiert}, B., {et~al.} 2013,
  \aap, 554, A58, \dodoi{10.1051/0004-6361/201220669}

\bibitem[{{S{\'a}nchez} {et~al.}(2016){S{\'a}nchez}, {P{\'e}rez},
  {S{\'a}nchez-Bl{\'a}zquez}, {Garc{\'\i}a-Benito}, {Ibarra-Mede},
  {Gonz{\'a}lez}, {Rosales-Ortega}, {S{\'a}nchez-Menguiano}, {Ascasibar},
  {Bitsakis}, {Law}, {Cano-D{\'\i}az}, {L{\'o}pez-Cob{\'a}}, {Marino}, {Gil de
  Paz}, {L{\'o}pez-S{\'a}nchez}, {Barrera-Ballesteros}, {Galbany}, {Mast},
  {Abril-Melgarejo}, \& {Roman-Lopes}}]{Sanchez2016}
{S{\'a}nchez}, S.~F., {P{\'e}rez}, E., {S{\'a}nchez-Bl{\'a}zquez}, P., {et~al.}
  2016, \rmxaa, 52, 171.
\newblock \doarXiv{1602.01830}

\bibitem[{{S{\'a}nchez} {et~al.}(2018){S{\'a}nchez}, {Avila-Reese},
  {Hernandez-Toledo}, {Cortes-Su{\'a}rez}, {Rodr{\'\i}guez-Puebla},
  {Ibarra-Medel}, {Cano-D{\'\i}az}, {Barrera-Ballesteros}, {Negrete},
  {Calette}, {de Lorenzo-C{\'a}ceres}, {Ortega-Minakata}, {Aquino},
  {Valenzuela}, {Clemente}, {Storchi-Bergmann}, {Riffel}, {Schimoia}, {Riffel},
  {Rembold}, {Brownstein}, {Pan}, {Yates}, {Mallmann}, \&
  {Bitsakis}}]{Sanchez2018}
{S{\'a}nchez}, S.~F., {Avila-Reese}, V., {Hernandez-Toledo}, H., {et~al.} 2018,
  \rmxaa, 54, 217.
\newblock \doarXiv{1709.05438}

\bibitem[{{S{\'a}nchez} {et~al.}(2022){S{\'a}nchez}, {Barrera-Ballesteros},
  {Lacerda}, {Mej{\'\i}a-Narvaez}, {Camps-Fari{\~n}a}, {Bruzual},
  {Espinosa-Ponce}, {Rodr{\'\i}guez-Puebla}, {Calette}, {Ibarra-Medel},
  {Avila-Reese}, {Hernandez-Toledo}, {Bershady}, {Cano-Diaz}, \&
  {Munguia-Cordova}}]{Sanchez2022}
{S{\'a}nchez}, S.~F., {Barrera-Ballesteros}, J.~K., {Lacerda}, E., {et~al.}
  2022, \apjs, 262, 36, \dodoi{10.3847/1538-4365/ac7b8f}

\bibitem[{{Scudder} {et~al.}(2012){Scudder}, {Ellison}, {Torrey}, {Patton}, \&
  {Mendel}}]{Scudder2012}
{Scudder}, J.~M., {Ellison}, S.~L., {Torrey}, P., {Patton}, D.~R., \& {Mendel},
  J.~T. 2012, \mnras, 426, 549, \dodoi{10.1111/j.1365-2966.2012.21749.x}

\bibitem[{{Tacchella} {et~al.}(2016){Tacchella}, {Dekel}, {Carollo},
  {Ceverino}, {DeGraf}, {Lapiner}, {Mandelker}, \& {Primack
  Joel}}]{Tacchella2016}
{Tacchella}, S., {Dekel}, A., {Carollo}, C.~M., {et~al.} 2016, \mnras, 457,
  2790, \dodoi{10.1093/mnras/stw131}

\bibitem[{{Tacchella} {et~al.}(2022){Tacchella}, {Conroy}, {Faber}, {Johnson},
  {Leja}, {Barro}, {Cunningham}, {Deason}, {Guhathakurta}, {Guo}, {Hernquist},
  {Koo}, {McKinnon}, {Rockosi}, {Speagle}, {van Dokkum}, \&
  {Yesuf}}]{Tacchella2022}
{Tacchella}, S., {Conroy}, C., {Faber}, S.~M., {et~al.} 2022, \apj, 926, 134,
  \dodoi{10.3847/1538-4357/ac449b}

\bibitem[{{Thomas} {et~al.}(2005){Thomas}, {Maraston}, {Bender}, \& {Mendes de
  Oliveira}}]{Thomas2005}
{Thomas}, D., {Maraston}, C., {Bender}, R., \& {Mendes de Oliveira}, C. 2005,
  \apj, 621, 673, \dodoi{10.1086/426932}

\bibitem[{{Tremonti} {et~al.}(2004){Tremonti}, {Heckman}, {Kauffmann},
  {Brinchmann}, {Charlot}, {White}, {Seibert}, {Peng}, {Schlegel}, {Uomoto},
  {Fukugita}, \& {Brinkmann}}]{Tremonti2004}
{Tremonti}, C.~A., {Heckman}, T.~M., {Kauffmann}, G., {et~al.} 2004, \apj, 613,
  898, \dodoi{10.1086/423264}

\bibitem[{{V{\'a}zquez} \& {Leitherer}(2005)}]{Vazquez2005}
{V{\'a}zquez}, G.~A., \& {Leitherer}, C. 2005, \apj, 621, 695,
  \dodoi{10.1086/427866}

\bibitem[{{Westfall} {et~al.}(2019){Westfall}, {Cappellari}, {Bershady},
  {Bundy}, {Belfiore}, {Ji}, {Law}, {Schaefer}, {Shetty}, {Tremonti}, {Yan},
  {Andrews}, {Brownstein}, {Cherinka}, {Coccato}, {Drory}, {Maraston},
  {Parikh}, {S{\'a}nchez-Gallego}, {Thomas}, {Weijmans}, {Barrera-Ballesteros},
  {Du}, {Goddard}, {Li}, {Masters}, {Ibarra Medel}, {S{\'a}nchez}, {Yang},
  {Zheng}, \& {Zhou}}]{Westfall2019}
{Westfall}, K.~B., {Cappellari}, M., {Bershady}, M.~A., {et~al.} 2019, \aj,
  158, 231, \dodoi{10.3847/1538-3881/ab44a2}

\bibitem[{{Willett} {et~al.}(2013){Willett}, {Lintott}, {Bamford}, {Masters},
  {Simmons}, {Casteels}, {Edmondson}, {Fortson}, {Kaviraj}, {Keel}, {Melvin},
  {Nichol}, {Raddick}, {Schawinski}, {Simpson}, {Skibba}, {Smith}, \&
  {Thomas}}]{Willett2013}
{Willett}, K.~W., {Lintott}, C.~J., {Bamford}, S.~P., {et~al.} 2013, \mnras,
  435, 2835, \dodoi{10.1093/mnras/stt1458}

\bibitem[{{Woo} \& {Ellison}(2019)}]{Woo2019}
{Woo}, J., \& {Ellison}, S.~L. 2019, \mnras, 487, 1927,
  \dodoi{10.1093/mnras/stz1377}

\bibitem[{{Yan} {et~al.}(2019){Yan}, {Chen}, {Lazarz}, {Bizyaev}, {Maraston},
  {Stringfellow}, {McCarthy}, {Meneses-Goytia}, {Law}, {Thomas}, {Falcon
  Barroso}, {S{\'a}nchez-Gallego}, {Schlafly}, {Zheng}, {Argudo-Fern{\'a}ndez},
  {Beaton}, {Beers}, {Bershady}, {Blanton}, {Brownstein}, {Bundy}, {Chambers},
  {Cherinka}, {De Lee}, {Drory}, {Galbany}, {Holtzman}, {Imig}, {Kaiser},
  {Kinemuchi}, {Liu}, {Luo}, {Magnier}, {Majewski}, {Nair}, {Oravetz},
  {Oravetz}, {Pan}, {Sobeck}, {Stassun}, {Talbot}, {Tremonti}, {Waters},
  {Weijmans}, {Wilhelm}, {Zasowski}, {Zhao}, \& {Zhao}}]{Yan2019}
{Yan}, R., {Chen}, Y., {Lazarz}, D., {et~al.} 2019, \apj, 883, 175,
  \dodoi{10.3847/1538-4357/ab3ebc}

\bibitem[{{Yu} {et~al.}(2022){Yu}, {Kalinova}, {Colombo}, {Bolatto}, {Wong},
  {Levy}, {Villanueva}, {S{\'a}nchez}, {Ho}, {Vogel}, {Teuben}, \&
  {Rubio}}]{Yu2022}
{Yu}, S.-Y., {Kalinova}, V., {Colombo}, D., {et~al.} 2022, \aap, 666, A175,
  \dodoi{10.1051/0004-6361/202244306}

\bibitem[{{Zhang} {et~al.}(2022){Zhang}, {Li}, {Leja}, {Whitaker}, {Nersesian},
  {Bezanson}, \& {van der Wel}}]{Zhang2022}
{Zhang}, J., {Li}, Y., {Leja}, J., {et~al.} 2022, arXiv e-prints,
  arXiv:2211.10450, \dodoi{10.48550/arXiv.2211.10450}

\bibitem[{{Zolotov} {et~al.}(2015){Zolotov}, {Dekel}, {Mandelker}, {Tweed},
  {Inoue}, {DeGraf}, {Ceverino}, {Primack}, {Barro}, \& {Faber}}]{Zolotov2015}
{Zolotov}, A., {Dekel}, A., {Mandelker}, N., {et~al.} 2015, \mnras, 450, 2327,
  \dodoi{10.1093/mnras/stv740}

\end{thebibliography}
\bibliographystyle{aasjournal}

\end{document}